\documentclass[12pt,twoside, a4paper]{article}
\pdfoutput=1

\def\mc{\mathcal}

\def\Slash{{\!\!\!\!/}}

\usepackage[dvips]{graphicx}
\usepackage{amssymb}
\usepackage{amssymb,amsmath}
\usepackage{graphicx}
\usepackage{dsfont}
\usepackage{caption}
\usepackage{subcaption}
\usepackage{mathtools}
\usepackage{verbatim}
\usepackage{graphicx}
\usepackage{multirow}
\usepackage[outline]{contour}
\usepackage{xcolor,colortbl}
\usepackage{pdflscape}
\input{epsf.sty}  
\contourlength{.1pt}
\graphicspath{ {./Figures/} }
\numberwithin{equation}{section}

\begin{document}
\begin{center}
\Large{\textbf{M5-branes wrapped on topological disks from maximal 7D gauged supergravity}}
\end{center}
\vspace{1 cm}
\begin{center}
\large{\textbf{Patharadanai Nuchino}$^a$ and \textbf{Parinya Karndumri}$^b$}
\end{center}
\begin{center}
$^a$ Quantum and Gravity Theory Research Group, Department of Physics, Faculty of Science, Ramkhamhaeng University, 282 Ramkhamhaeng Road, Bang Kapi, Bangkok 10240, Thailand\\
E-mail: danai.nuchino@hotmail.com \\
$^b$ String Theory and Supergravity Group, Department
of Physics, Faculty of Science, Chulalongkorn University, 254 Phayathai Road, Pathumwan, Bangkok 10330, Thailand\\
E-mail: parinya.ka@hotmail.com \vspace{1 cm} \\
\end{center}
\begin{abstract}
We study new supersymmetric $AdS_5 \times \Sigma$ solutions for $\Sigma$ being a topological disk from the maximal gauged supergravity in seven dimensions with $CSO(p,q,5-p-q)$ gauge groups. These gauged supergravities can be obtained from consistent truncations of eleven-dimensional supergravity on $H^{p,q} \circ T^{5-p-q}$. We find a number of solutions with $SO(2)\times SO(2)$ and $SO(2)\times SO(3)$ symmetries from $SO(5)$ and $SO(3,2)$ gauge groups as well as $SO(2)_{\textrm{diag}}$ symmetric solutions from $SO(5)$ and $SO(4,1)$ gauge groups. Solutions with $SO(2)\times SO(3)$ symmetry preserve $16$ supercharges while the other solutions preserve only $8$ supercharges. Although some of the solutions give a holographic description of the dual $N=1$ SCFTs in four dimensions with finite central charges, most of the solutions describe codimension-2 defects within $N=(2,0)$ theories in six dimensions. For solutions with $SO(2)\times SO(2)$ and $SO(2)_{\textrm{diag}}$ symmetries from $SO(5)$ gauge group for which a complete truncation ansatz of eleven-dimensional supergravity on $S^4$ is known, we also uplift the solutions to eleven dimensions and compute the holographic central charge of the dual four-dimensional $N=1$ SCFT with $SO(2)_{\textrm{diag}}$ symmetry. 
\end{abstract}
\newpage
\tableofcontents

\section{Introduction}
Recently, new mechanisms for preserving supersymmetry have attracted great interest in supersymmetric $AdS_n\times \Sigma$ solutions for $\Sigma$ being a two-dimensional orbifold with non-constant curvature. The two-dimensional space $\Sigma$ can be a spindle \cite{Ferro_D3,Gaunlett_super_spindle} or a topological disk \cite{Bah_AD,Bah_M5} with two or one orbifold singularities, respectively. Topologically, a spindle is a two-sphere with orbifold singularities at the two poles while a topological disk has one orbifold singularity at the center with a non-trivial $U(1)$ holonomy on the boundary. These $AdS_n\times \Sigma$ solutions describe near horizon geometries of $n$-branes wrapped on $\Sigma$ which, according to the AdS/CFT correspondence \cite{maldacena,Gubser_AdS_CFT,Witten_AdS_CFT}, lead to new superconformal field theories (SCFTs) in $n-1$ dimensions arising from $(n+1)$-dimensional SCFTs, dual to the $AdS_{n+2}$ vacua, compactified on a spindle or a topological disk. The construction along these lines have led to a number of new $AdS_n\times \Sigma$ solutions from gauged supergravities in various dimensions \cite{Hosseini_D3}-\cite{6D_F4_disk}, see also a generalization to higher-dimensional orbifolds in \cite{higher_D_orbifold1}-\cite{M5_WCP2}. 
\\
\indent In this work, we are interested in supersymmetric $AdS_5\times \Sigma$ solutions with $\Sigma$ being a topological disk from maximal $N=4$ gauged supergravity in seven dimensions. The original $AdS_5\times \Sigma$ solution has been found from a $U(1)\times U(1)$ truncation of $N=4$ gauged supergravity with $SO(5)$ gauge group \cite{Bah_AD,Bah_M5}. In this solution, one of the two $U(1)$ gauge fields vanishes and the two $U(1)\times U(1)$ invariant scalars are related leading to an enhanced $U(1)\times SU(2)\sim SO(2)\times SO(3)$ symmetry. The solution perserves half of the original supersymmetry and has been shown to give a holographic dual of an $N=2$ SCFT of Argyres-Douglas type. The solution has been generalized to a general choice of regular punctures in \cite{M5_disk_Couzen1}, and a more general solution with one more scalar added has been considered in \cite{M5_disk_Couzen2}. All of these solutions give rise to holographic descriptions of the dual $N=2$ SCFTs in four dimensions. In the present paper, we will extend these results by considering $AdS_5\times \Sigma$ solutions preserving eight supercharges dual to $N=1$ four-dimensional SCFTs. As pointed out in \cite{M5_disk_Couzen2}, there do not exist $N=1$ $AdS_5\times \Sigma$ solutions with $\Sigma$ being a spindle, so only solutions with $\Sigma$ being a topological disk are possible. However, these $N=1$ $AdS_5\times \Sigma$ solutions have not been considered in \cite{M5_disk_Couzen2}. We will study these solutions in the present paper.       
\\
\indent Moreover, unlike all the previous results, we will study supersymmetric $AdS_5\times \Sigma$ solutions in seven-dimensional $N=4$ gauged supergravity with non-compact and non-semisimple $CSO(p,q,5-p-q)$ gauge group constructed in \cite{N4_7D_Henning}, see also \cite{7D_Pernici} and \cite{Noncompact_7D} for earlier constructions. These gauge groups are obtained from the embedding tensor in $\mathbf{15}$ representation of the global symmetry $SL(5)$ and have been shown to arise from consistent truncations of eleven-dimensional supergravity on $H^{p,q}\circ T^{5-p-q}$ \cite{Henning_Hohm1, Henning_Emanuel}. The manifold $H^{p,q}\circ T^r$ is a product of a $(p+q-1)$-dimensional hyperbolic space and an $r$-torus. The former is in general non-compact except for $p=0$ or $q=0$ in which $H^{p,0}$ and $H^{0,q}$ become a $(p-1)$-sphere and a $(q-1)$-sphere, respectively. With some appropriate discrete quotient, it should be possible to make the $H^{p,q}$ part into a compact space. However, the complete truncation ansatz has appeared only for the case of $S^4$ \cite{M-theory_on_S4_1,M-theory_on_S4_2}. Accordingly, in this work, we will uplift only solutions from $SO(5)$ gauge group corresponding to a truncation of eleven-dimensional supergravity on $S^4$ although, in principle, all solutions found here can be embedded in eleven dimensions. 
\\
\indent The paper is organized as follows. In section \ref{N4_7D_SUGRA}, we briefly review the maximal gauged supergravity in seven dimensions with $CSO(p,q,5-p-q)$ gauge groups. Supersymmetric $AdS_5\times\Sigma$ solutions with $SO(2)\times SO(2)$, $SO(2)_{\textrm{diag}}$, and $SO(2)\times SO(3)$ symmetries are given in sections \ref{Y_SO(2)xSO(2)_Sec}, \ref{Y_SO(2)D_Sec}, and \ref{Y_SO(2)_Sec}, respectively. Some conclusions and comments are given in section \ref{conclusion}. In appendix \ref{App_field_eqs}, we list all the bosonic field equations for the maximal seven-dimensional gauged supergravity under consideration here.
\section{$N=4$ gauged supergravity in seven dimensions}\label{N4_7D_SUGRA}
We first give a brief review of seven-dimensional $N = 4$ gauged supergravity in the embedding tensor formalism constructed in \cite{N4_7D_Henning}. We mainly collect relevant formulae involving the bosonic Lagrangian and fermionic supersymmetry transformations which are essential for finding supersymmetric solutions and refer to \cite{N4_7D_Henning} for more details.
\\
\indent The $N=4$ supergravity multiplet in seven dimensions consists of the following field content
\begin{equation}\label{fieldcon}
(e^{\hat{\mu}}_\mu, \psi^a_\mu, A^{MN}_\mu, B_{\mu\nu M}, \chi^{abc}, {\mathcal{V}_M}^A).
\end{equation}
The component fields are given by the graviton $e^{\hat{\mu}}_\mu$, four gravitini $\psi^a_\mu$, ten vectors $A^{MN}_\mu=A^{[MN]}_\mu$, five two-form fields $B_{\mu\nu M}$, sixteen spin-$\frac{1}{2}$ fermions $\chi^{abc}=\chi^{[ab]c}$ and fourteen scalar fields described by the $SL(5)/SO(5)$ coset representative ${\mathcal{V}_M}^A$. In this paper, curved and flat space-time indices are denoted by $\mu,\nu,\ldots$ and $\hat{\mu},\hat{\nu},\ldots$, respectively. Lower (upper) $M,N=1,...,5$ indices refer to the (anti-) fundamental representation \textbf{5} ($\overline{\mathbf{5}}$) of the global $SL(5)$ symmetry. 
\\
\indent Fermionic fields are described by symplectic Majorana spinors subject to the conditions
\begin{equation}
\bar{\psi}^T_{\mu a}=\Omega_{ab}C\psi_\mu^b\qquad \textrm{and}\qquad \bar{\chi}^T_{abc}=\Omega_{ad}\Omega_{be}\Omega_{cf}C\chi^{def}
\end{equation}
where $C$ denotes the charge conjugation matrix obeying
\begin{equation}
C=C^T=-C^{-1}=-C^\dagger\, .
\end{equation}
$\Omega_{ab}=\Omega_{[ab]}$ is the $USp(4)$ symplectic form satisfying
\begin{equation}
(\Omega_{ab})^*=\Omega^{ab}\qquad \textrm{and}\qquad \Omega_{ac}\Omega^{bc}=\delta^b_a\, .
\end{equation}
The fermionic fields transform under the local $SO(5)\sim USp(4)$ R-symmetry as the fundamental representation of $USp(4)$ or spinor representation of $SO(5)$ denoted by indices $a,b,\ldots =1,\ldots ,4$. Accordingly, the four gravitini $\psi^a_\mu$ and the spin-$\frac{1}{2}$ fields $\chi^{abc}$ transform as $\mathbf{4}$ and $\mathbf{16}$, respectively. In particular, $\chi^{abc}$ satisfy the following conditions
\begin{equation}
\chi^{[abc]}=0\qquad \textrm{and}\qquad \Omega_{ab}\chi^{abc}=0\, .
\end{equation}
We also recall that raising and lowering of $USp(4)$ indices by $\Omega^{ab}$ and $\Omega_{ab}$ correspond to complex conjugation.
\\
\indent The fourteen scalars are described by the $SL(5)/SO(5)$ coset representative ${\mc{V}_M}^A$ transforming under the global $SL(5)$ and local $SO(5)$ symmetries by left and right multiplications. Indices $M=1,2,\ldots, 5$ and $A=1,2,\ldots ,5$ are accordingly $SL(5)$ and $SO(5)$ fundamental indices, respectively. The $SO(5)$ vector indices of ${\mc{V}_M}^A$ can also be rewritten in terms of a pair of antisymmetric $USp(4)$ fundamental indices as ${\mc{V}_M}^{ab}={\mc{V}_M}^{[ab]}$ satisfying the relation
\begin{equation}
{\mathcal{V}_M}^{ab}\Omega_{ab}=0\, .
\end{equation}
Similarly, the inverse of ${\mathcal{V}_M}^A$ denoted by ${\mc{V}_A}^M$ can be written as ${\mathcal{V}_{ab}}^M$ with 
\begin{equation}
{\mathcal{V}_M}^{ab}{\mathcal{V}_{ab}}^N=\delta^N_M \qquad \textrm{and} \qquad {\mathcal{V}_{ab}}^M{\mathcal{V}_M}^{cd}=\delta^{[c}_{a}\delta^{d]}_{b}-\frac{1}{4}\Omega_{ab}\Omega^{cd}\, .
\end{equation}
Explicitly, both ${\mc{V}_M}^{ab}$ and ${\mc{V}_{ab}}^M$ can be written in terms of $SO(5)$ gamma matrices $\Gamma_A$ as
\begin{equation}
{\mathcal{V}_M}^{ab}=\frac{1}{2} {\mathcal{V}_M}^A(\Gamma_A)^{ab}\qquad \textrm{and}\qquad
{\mathcal{V}_{ab}}^M=\frac{1}{2} {\mathcal{V}_A}^M(\Gamma^A)_{ab}\, .
\end{equation}
The matrices $\Gamma_A$ satisfy the following relations
\begin{eqnarray}
\{\Gamma_A,\Gamma_B\}&=&2\delta_{AB}\mathds{1}_4,\qquad  (\Gamma_A)^{ab}=-(\Gamma_A)^{ba}, \nonumber \\
\Omega_{ab}(\Gamma_A)^{ab}&=&0, \qquad ((\Gamma_A)^{ab})^*=\Omega_{ac}\Omega_{bd}(\Gamma_A)^{cd}
\end{eqnarray}
in which $\mathds{1}_n$ is an $n\times n$ identity matrix. 
\\
\indent The bosonic Lagrangian of the $N=4$ seven-dimensional gauged supergravity can be written as
\begin{eqnarray}
e^{-1}\mathcal{L}&=&\frac{1}{2}R-\mathcal{M}_{MP}\mathcal{M}_{NQ}\mathcal{H}_{\mu\nu}^{(2)MN}\mathcal{H}^{(2)PQ\mu\nu }-\frac{1}{6}\mathcal{M}^{MN}\mathcal{H}_{\mu\nu\rho M}^{(3)}{\mathcal{H}^{(3)\mu\nu\rho}}_ N\nonumber \\
& &+\frac{1}{8}(D_\mu\mathcal{M}_{MN})(D^\mu\mathcal{M}^{MN})-e^{-1}\mathcal{L}_{VT}-\mathbf{V}\label{BosLag}
\end{eqnarray}
while the supersymmetry transformations of fermions read
\begin{eqnarray}
\delta\psi^a_\mu&=& D_\mu\epsilon^a-g\gamma_\mu A^{ab}_1\Omega_{bc}\epsilon^c+\frac{1}{15}\mathcal{H}_{\nu\rho\lambda M}^{(3)}({\gamma_\mu}^{\nu\rho\lambda}-\frac{9}{2}\delta^\nu_\mu\gamma^{\rho\lambda})\Omega^{ab}{\mathcal{V}_{bc}}^M\epsilon^c\nonumber \\ & &+\frac{1}{5}\mathcal{H}_{\nu\rho}^{(2)MN}({\gamma_\mu}^{\nu\rho}-8\delta^\nu_\mu\gamma^{\rho}){\mathcal{V}_M}^{ad}\Omega_{de}{\mathcal{V}_N}^{eb}\Omega_{bc}\epsilon^c,\\
\delta\chi^{abc}&=& 2\Omega^{cd}{P_{\mu de}}^{ab}\gamma^\mu\epsilon^e+gA^{d,abc}_2\Omega_{de}\epsilon^e\nonumber \\ & &
+2\mathcal{H}_{\mu\nu}^{(2)MN}\gamma^{\mu\nu}\Omega_{de}\left[{\mathcal{V}_M}^{cd}{\mathcal{V}_N}^{e[a}\epsilon^{b]}-\frac{1}{5}(\Omega^{ab}\delta^c_g-\Omega^{c[a}\delta^{b]}_g){\mathcal{V}_M}^{gf}\Omega_{fh}{\mathcal{V}_N}^{hd}\epsilon^{e}\right] \nonumber \\ & &
-\frac{1}{6}\mathcal{H}_{\mu\nu\rho M}^{(3)}\gamma^{\mu\nu\rho}{\mathcal{V}_{fe}}^M\left[\Omega^{af}\Omega^{be}\epsilon^c-\frac{1}{5}(\Omega^{ab}\Omega^{cf}+4\Omega^{c[a}\Omega^{b]f})\epsilon^e\right].
\end{eqnarray}
The covariant derivative of the supersymmetry parameters is defined by
\begin{equation}
D_\mu\epsilon^a=\nabla_\mu\epsilon^a-{Q_{\mu b}}^a\epsilon^b
\end{equation}
with $\nabla_\mu$ being the space-time covariant derivative. The composite connection ${Q_{\mu a}}^b$ and the vielbein on the $SL(5)/SO(5)$ coset ${P_{\mu ab}}^{cd}$ are obtained from
\begin{equation}
{P_{\mu ab}}^{cd}+2{Q_{\mu [a}}^{[c}\delta^{d]}_{b]}= {\mathcal{V}_{ab}}^M(\partial_\mu{\mathcal{V}_M}^{cd}-gA^{PQ}_\mu{X_{PQ,M}}^N{\mathcal{V}_N}^{cd}).
\end{equation}
\indent A general gauging can be described by an $SL(5)$ covariant tensor called the embedding tensor. Supersymmetry implies that the embedding tensor can have only two components parametrized by $Y_{MN}$ and $Z^{MN,P}$ tensors with $Y_{MN}=Y_{(MN)}$, $Z^{MN,P}=Z^{[MN],P}$ and $Z^{[MN,P]}=0$ corresponding to representations $\mathbf{15}$ and $\overline{\mathbf{40}}$ of $SL(5)$, respectively. As a result, the gauge generators in the fundamental representation $\mathbf{5}$ of $SL(5)$ can be written in term of the embedding tensor as
\begin{equation}
{X_{MN,P}}^Q= {\Theta_{MN,P}}^Q=\delta^Q_{[M}Y_{N]P}-2\epsilon_{MNPRS}Z^{RS,Q}\, .
\end{equation}
In addition, closure of the gauge algebra also requires the quadratic constraint
\begin{equation}
Y_{MQ}Z^{QN,P}+2\epsilon_{MRSTU}Z^{RS,N}Z^{TU,P}=0\, .
\end{equation}
\indent The fermion shift matrices $A_1$ and $A_2$ are given by
\begin{eqnarray}
A^{ab}_1&=& -\frac{1}{4\sqrt{2}}\left(\frac{1}{4}B\Omega^{ab}+\frac{1}{5}C^{ab}\right),\label{A1}\\
A^{d,abc}_2&=&\frac{1}{2\sqrt{2}}\left[\Omega^{ec}\Omega^{fd}({C^{ab}}_{ef}-{B^{ab}}_{ef})+\frac{1}{4}(C^{ab}\Omega^{cd}+\frac{1}{5}\Omega^{ab}C^{cd}+\frac{4}{5}\Omega^{c[a}C^{b]d})\right]\nonumber\\\label{A2}
\end{eqnarray}
with 
\begin{eqnarray}\label{BCfunctions}
B&=&\frac{\sqrt{2}}{5}\Omega^{ac}\Omega^{bd}Y_{ab,cd},\\
{{B^{ab}}_{cd}}&=&\sqrt{2}\left[\Omega^{ae}\Omega^{bf}\delta^{[g}_c \delta^{h]}_d-\frac{1}{5}(\delta^{[a}_c \delta^{b]}_d-\frac{1}{4}\Omega^{ab}\Omega_{cd})\Omega^{eg}\Omega^{fh}\right]Y_{ef,gh},\\
C^{ab}&=&8\Omega_{cd}Z^{(ac)[bd]},\\
{C^{ab}}_{cd}&=&8\left(-\Omega_{ce}\Omega_{df}\delta^{[a}_g \delta^{b]}_h+\Omega_{g(c}\delta^{[a}_{d)} \delta^{b]}_e\Omega_{fh}\right)Z^{(ef)[gh]}\, .
\end{eqnarray}
The ``dressed'' components of the embedding tensor are defined by
\begin{eqnarray}
Y_{ab,cd}&=&{\mathcal{V}_{ab}}^M{\mathcal{V}_{cd}}^NY_{MN},\\
\textrm{and}\qquad Z^{(ac)[ef]}&=&\sqrt{2}{\mathcal{V}_M}^{ab}{\mathcal{V}_N}^{cd}{\mathcal{V}_P}^{ef}\Omega_{bd}Z^{MN,P}\, .
\end{eqnarray}
\indent A unimodular symmetric matrix $\mc{M}_{MN}$ is defined by
\begin{equation}\label{fullM}
\mathcal{M}_{MN}={\mathcal{V}_M}^{ab}{\mathcal{V}_N}^{cd}\Omega_{ac}\Omega_{bd}
\end{equation}
together with its inverse
\begin{equation}
\mathcal{M}^{MN}={\mathcal{V}_{ab}}^M{\mathcal{V}_{cd}}^N\Omega^{ac}\Omega^{bd}\, .
\end{equation}
The scalar potential is given by
\begin{eqnarray}\label{scalarPot}
\mathbf{V}&=& \frac{g^2}{64}\left[2\mathcal{M}^{MN}Y_{NP}\mathcal{M}^{PQ}Y_{QM}-(\mathcal{M}^{MN}Y_{MN})^2\right]\nonumber \\
& &+g^2Z^{MN,P}Z^{QR,S}\left(\mathcal{M}_{MQ}\mathcal{M}_{NR}\mathcal{M}_{PS}-\mathcal{M}_{MQ}\mathcal{M}_{NP}\mathcal{M}_{RS}\right)
\nonumber \\
&=&-15A_1^{ab}A_{1 ab}+ \frac{1}{8}A_2^{a,bcd}A_{2 a,bcd}
\end{eqnarray}
Finally, $\mathcal{L}_{VT}$ denotes the topological term describing the interaction between the vector and tensor fields. The explcit form of this term can be found in \cite{N4_7D_Henning}.  
\\
\indent The gauge covariant field strength tensors of vector and two-form fields are defined by
\begin{eqnarray}
\mathcal{H}_{\mu\nu}^{(2)MN}&=&F^{MN}_{\mu\nu}+gZ^{MN,P}B_{\mu\nu P},\label{ModTensor1}\\
\mathcal{H}_{\mu\nu\rho M}^{(3)}&=&gY_{MN}S^N_{\mu\nu\rho}+3D_{[\mu}B_{\nu\rho]M}\nonumber\\
&&+6\epsilon_{MNPQR}A^{NP}_{[\mu}(\partial_\nu A^{QR}_{\rho]}+\frac{2}{3}g {X_{ST,U}}^QA^{RU}_\nu A^{ST}_{\rho]})\label{ModTensor2}
\end{eqnarray}
with the usual non-abelian gauge field strength
\begin{equation}\label{Ful2Form}
F^{MN}_{\mu\nu}=2\partial_{[\mu}A^{MN}_{\nu]}+g{(X_{PQ})_{RS}}^{MN}A^{PQ}_{[\mu}A^{RS}_{\nu]}
\end{equation}
where ${(X_{MN})_{PQ}}^{RS}=2{X_{MN,[P}}^{[R}\delta^{S]}_{Q]}$. These field strengths satisfy the following Bianchi identities
\begin{eqnarray}
D_{[\mu}\mathcal{H}^{(2)MN}_{\nu\rho]}&=&\frac{1}{3}gZ^{MN,P}\mathcal{H}^{(3)}_{\mu\nu\rho P}, \label{DefBianchi1}\\
D_{[\mu}\mathcal{H}^{(3)}_{\nu\rho\lambda] M}&=&\frac{3}{2}\epsilon_{MNPQR}\mathcal{H}^{(2)NP}_{[\mu\nu}\mathcal{H}^{(2)QR}_{\rho\lambda]}+\frac{1}{4}gY_{MN}\mathcal{H}^{(4)N}_{\mu\nu\rho\lambda}\label{DefBianchi2}
\end{eqnarray}
where the covariant field strengths of the three-form fields are defined by
\begin{eqnarray}\label{4-form}
Y_{MN}\mathcal{H}^{(4)N}_{\mu\nu\rho\lambda}&=&Y_{MN}\left[4D_{[\mu}S^N_{\nu\rho\lambda]}+6F^{NP}_{[\mu\nu}B_{\rho\lambda] P}+3gZ^{NP,Q}B_{[\mu\nu P}B_{\rho\lambda] Q}\right.\nonumber \\
& &+4g\epsilon_{PQRVW}{X_{ST,U}}^VA^{NP}_{[\mu}A^{QR}_\nu A^{ST}_\rho A^{UW}_{\lambda]}\nonumber \\
& &\left.+8\epsilon_{PQRST}A^{NP}_{[\mu}A^{QR}_\nu\partial_\rho A^{ST}_{\lambda]}\right].
\end{eqnarray}
\indent In this paper, we are interested in the maximal gauged supergravity with $CSO(p,q,5-p-q)$ gauge groups obtained from the embedding tensor $Y_{MN}$ in representation $\mathbf{15}$ of the global symmetry $SL(5)$. With $Z^{MN,P}=0$, the corresponding gauge generators are simply given by
\begin{equation}
{(X_{MN})_P}^Q=\delta^Q_{[M}Y_{N]P}
\end{equation}
for any symmetric tensor $Y_{MN}$. We can also use the $SL(5)$ symmetry to fix the form of $Y_{MN}$ to be
\begin{equation}
Y_{MN}=\text{diag}(\underbrace{1,..,1}_p,\underbrace{-1,..,-1}_q,\underbrace{0,..,0}_r),\qquad p+q+r=5\, .
\end{equation}

\section{$SO(2)\times SO(2)$ symmetric solutions}\label{Y_SO(2)xSO(2)_Sec} 
We begin with supersymmetric $AdS_5\times \Sigma$ solutions with $SO(2)\times SO(2)$ symmetry. We will choose the generators of this $SO(2)\times SO(2)$ subgroup to be $X_{12}$ and $X_{34}$. There are many gauge groups containing this $SO(2)\times SO(2)$ subgroup. All of these gauge groups can be collectively described by the embedding tensor of the form
\begin{equation}\label{SO(2)xSO(2)Ytensor}
Y_{MN}=\text{diag}(+1,+1,\kappa,\kappa,\lambda).
\end{equation}
This corresponds to $SO(5)$ ($\kappa=\lambda=1$), $SO(4,1)$ ($\kappa=-\lambda=1$), $SO(3,2)$ ($\kappa=-\lambda=-1$), $CSO(4,0,1)$ ($\kappa=1$, $\lambda=0$), and $CSO(2,2,1)$ ($\kappa=-1$, $\lambda=0$) gauge groups. 
\\
\indent Among the fourteen scalars in $SL(5)/SO(5)$ coset, there are two $SO(2)\times SO(2)$ singlets corresponding to the following $SL(5)$ non-compact generators 
\begin{equation}
\hat{Y}_1=e_{1,1}+e_{2,2}-2e_{5,5}\qquad \textrm{and}\qquad
\hat{Y}_2=e_{3,3}+e_{4,4}-2e_{5,5}\label{Y12_def}
\end{equation}
in which we have introduced $GL(5)$ matrices defined by
\begin{equation}
(e_{IJ})_{MN}=\delta_{IM}\delta_{JN}\, .
\end{equation}
The corresponding $SL(5)/SO(5)$ coset representative is then given by 
\begin{equation}\label{YSO(2)xSO(2)Ys}
\mathcal{V}=e^{\phi_1\hat{Y}_1+\phi_2\hat{Y}_2}\, .
\end{equation}
The kinetic term for the non-vanishing scalar fields takes the form
\begin{equation}
\frac{1}{8}(D_\mu\mathcal{M}_{MN})(D^\mu\mathcal{M}^{MN})=-3(\partial_\mu\phi_1)^2-4\partial_\mu\phi_1\partial^\mu\phi'_2-3(\partial_\mu\phi_2)^2\, .
\end{equation}
It is also useful to define the scalar matrix $\mathbb{G}^{ij}$ by rewritting the kinetic term as $\mathbb{G}^{ij}\partial_\mu\phi_i\partial^\mu\phi_j$ with $i,j=1,2$. In particular, this leads to the inverse matrix of the form
\begin{equation}
\mathbb{G}_{ij}=-\frac{1}{10}\begin{pmatrix} 3 & -2 \\ -2 & 3\end{pmatrix}.
\end{equation}
This matrix will play an important role in the analysis of the BPS equations.
\\
\indent Following \cite{Bah_M5}, we take the ansatz for the seven-dimensional metric to be
\begin{equation}\label{7Dmetrix}
ds^2_7=f(r)ds^2_{AdS_5}+h_1(r)dr^2+h_2(r)dz^2
\end{equation}
where the metric on $AdS_5$ with unit radius is given by
\begin{equation}
ds^2_{AdS_5}=\frac{1}{\rho^2}(dx^2_{1,3}+d\rho^2)
\end{equation}
with $dx^2_{1,3}=\eta_{mn}dx^m dx^n$, $m,n= 0,\ldots, 3$ being the flat metric on the four-dimensional Minkowski space. $r$ and $z$ are respectively radial and angular coordinates on the topological disk $\Sigma$. The seven-dimensional curved and flat space-time indices will be split as $\mu=(m,\rho,r,z)$ and $\hat{\mu}=(\hat{m},\hat{\rho},\hat{r},\hat{z})$, respectively.
\\
\indent The ansatz for the $SO(2)\times SO(2)$ gauge fields is taken to be
\begin{equation}\label{SO(2)xSO(2)_Gaugefield}
A^{12}_z=\mathbb{A}_1(r)\qquad\text{ and }\qquad  A^{34}_z=\mathbb{A}_2(r)\, .
\end{equation}
Furthermore, we will set all the remaining tensor fields to zero and set $\phi_i=\phi_i(r)$. 

\subsection{Analysis of the BPS equations}
We will use the Killing spinors for the unbroken supersymmetry of the form
\begin{equation}\label{7DKilling}
\epsilon^a=n^a\,\vartheta\otimes\eta\, .
\end{equation}
$n^a$ are a constant object in the fundamental representation of $USp(4)$. $\eta$ is a two-component spinor on $\Sigma$ depending on $r$ and $z$ while $\vartheta$ is a four-component Killing spinor on $AdS_5$ satisfying
\begin{equation}\label{5DKilling}
\nabla_{\hat{\alpha}}^{AdS_5}\vartheta=\frac{s}{2}\mathbf{g}_{\hat{\alpha}}\vartheta
\end{equation}
with $\hat{\alpha}=(\hat{m},\hat{\rho})=0,1,2,3,4$ being a flat space-time index on $AdS_5$. The parameter $s=\pm1$ is an arbitrary sign, and $\mathbf{g}_{\hat{\alpha}}$ are five-dimensional $4\times4$ gamma matrices satisfying the Clifford algebra 
\begin{equation}
\{\mathbf{g}_{\hat{\alpha}},\mathbf{g}_{\hat{\beta}}\}=2\eta_{\hat{\alpha}\hat{\beta}} \qquad\textrm{with}\qquad \eta_{\hat{\alpha}\hat{\beta}}=\text{diag}(-++++).
\end{equation}
\indent With the following choice of vielbein one-forms
\begin{eqnarray}
& &e^{\hat{m}}_{(1)}=\frac{\sqrt{f(r)}}{\rho}dx^{m},\qquad e^{\hat{\rho}}_{(1)}=\frac{\sqrt{f(r)}}{\rho}d\rho,\nonumber \\
& & e^{\hat{r}}_{(1)}=\sqrt{h_1(r)}dr,\qquad\ \ e^{\hat{z}}_{(1)}=\sqrt{h_2(r)}dz, \label{7Dvielbein}
\end{eqnarray}
we can straightforwardly compute all non-vanishing components of the spin connection. These are given by
\begin{eqnarray}
& &\omega_{(1)}^{\hat{m}\hat{\rho}}=-\frac{1}{\sqrt{f}}e^{\hat{m}},\qquad\omega_{(1)}^{\hat{m}\hat{r}}=\frac{f'}{2f\sqrt{h_1}}e^{\hat{m}},\nonumber \\
& &\omega_{(1)}^{\hat{\rho}\hat{r}}=\frac{f'}{2f\sqrt{h_1}}e^{\hat{\rho}},\qquad\omega_{(1)}^{\hat{z}\hat{r}}=\frac{h'_2}{2h_2\sqrt{h_1}}e^{\hat{z}}\, .\label{spinCon}
\end{eqnarray}
From now on, we will use primes to denote $r$-derivatives and mostly suppress arguments of the $r$-dependent functions for convenience.
\\
\indent With $Z^{MN,P}=0$, the field strength tensors of the $SO(2)\times SO(2)$ gauge fields are given by
\begin{equation}\label{SO(2)xSO(2)_GaugeFS}
\mathcal{H}_{rz}^{(2)12}=F^{12}_{rz}=\mathbb{A}'_1\qquad\text{ and }\qquad  \mathcal{H}_{rz}^{(2)34}=F^{34}_{rz}=\mathbb{A}'_2.
\end{equation}
This ansatz immediately gives $\epsilon_{MNPQR}\mathcal{H}^{(2)NP}_{[\mu\nu}\mathcal{H}^{(2)QR}_{\rho\lambda]}=0$. Accordingly, from the Bianchi identities, we can consistently set $B_{\mu\nu M}=0$ and $S^M_{\mu\nu\rho}=0$ leading to vanishing $\mathcal{H}^{(3)}_{\mu\nu\rho P}$ and $\mathcal{H}^{(4)M}_{\mu\nu\rho\lambda}$. It can also be verified that all the the Bianchi identities given in \eqref{DefBianchi1} and \eqref{DefBianchi2} together with the field equations for the two-form and three-form fields \eqref{3formfieldEQ} and \eqref{fullSD} are all satisfied.
\\
\indent On the other hand, the field equation for the vector fields \eqref{H2_field_eqs} reduces to
\begin{eqnarray}
\mathbb{A}''_1&=&-\left(\frac{5f'}{f}-\frac{h'_1}{h_1}-\frac{h'_2}{h_2}+8\phi'_1\right)\frac{\mathbb{A}'_1}{2},\label{SO(2)xSO(2)_Diff_eq1}\\
\mathbb{A}''_2&=&-\left(\frac{5f'}{f}-\frac{h'_1}{h_1}-\frac{h'_2}{h_2}+8\phi'_2\right)\frac{\mathbb{A}'_2}{2}\, .\label{SO(2)xSO(2)_Diff_eq2}
\end{eqnarray}
The most general solution to these equations is given by 
\begin{equation}
\mathbb{A}'_1=\frac{a_1e^{-4\phi_1}\sqrt{h_1h_2}}{f^{5/2}}\qquad \textrm{and} \qquad \mathbb{A}'_2=\frac{a_2e^{-4\phi_2}\sqrt{h_1h_2}}{f^{5/2}}\label{Gen_SO(2)xSO(2)_Ansatz}
\end{equation}
where $a_1$ and $a_2$ are integration constants.
\\
\indent We now consider first-order BPS equations derived from setting the supersymmetry transformations of fermionic fields to zero. We begin with the variation of the gravitinos $\delta\psi_\mu^a=0$ along $m$, $\rho$, $r$, and $z$ directions. These are respectively given by
\begin{eqnarray}
0&=&2\partial_m\epsilon^a-\frac{1}{\sqrt{f}}\gamma_m\gamma^{\hat{\rho}}\epsilon^a+\frac{f'}{2f\sqrt{h_1}}\gamma_m\gamma^{\hat{r}}\epsilon^a-g\mathcal{W}\gamma_m\epsilon^a\nonumber\\
&&-\frac{2}{5\sqrt{h_1h_2}}\left[e^{2\phi_1}\mathbb{A}'_1{(\Gamma_{12})^a}_b+e^{2\phi_2}\mathbb{A}'_2{(\Gamma_{34})^a}_b\right]\gamma_m\gamma^{\hat{r}\hat{z}}\epsilon^b,\\
0&=&2\partial_\rho\epsilon^a+\frac{f'}{2f\sqrt{h_1}}\gamma_\rho\gamma^{\hat{r}}\epsilon^a-g\mathcal{W}\gamma_\rho\epsilon^a\nonumber\\
&&-\frac{2}{5\sqrt{h_1h_2}}\left[e^{2\phi_1}\mathbb{A}'_1{(\Gamma_{12})^a}_b+e^{2\phi_2}\mathbb{A}'_2{(\Gamma_{34})^a}_b\right]
\gamma_\rho\gamma^{\hat{r}\hat{z}}\epsilon^b,\\
0&=&2\partial_r\epsilon^a-g\mathcal{W}\gamma_r\epsilon^a+\frac{8}{5\sqrt{h_2}}\left[e^{2\phi_1}\mathbb{A}'_1{(\Gamma_{12})^a}_b+e^{2\phi_2}\mathbb{A}'_2{(\Gamma_{34})^a}_b\right]
\gamma^{\hat{z}}\epsilon^b,\\
0&=&2\partial_z\epsilon^a-g\left[\mathbb{A}_1{(\Gamma_{12})_b}^a+\kappa \mathbb{A}_2{(\Gamma_{34})_b}^a\right]\epsilon^b+\frac{h'_2}{2h_2\sqrt{h_1}}\gamma_z\gamma^{\hat{r}}\epsilon^a-g\mathcal{W}\gamma_z\epsilon^a\nonumber\\
&&-\frac{8}{5\sqrt{h_1}}\left[e^{2\phi_1}\mathbb{A}'_1{(\Gamma_{12})^a}_b+e^{2\phi_2}\mathbb{A}'_2{(\Gamma_{34})^a}_b\right]\gamma^{\hat{r}}\epsilon^b\label{last_raw_BPS}
\end{eqnarray}
where
\begin{equation}
\mathcal{W}=\frac{1}{40}\left(2e^{-2\phi_1}+2\kappa e^{-2\phi_2}+\lambda e^{4(\phi_1+\phi_2)}\right)
\end{equation}
obtained from the fermion-shift matrix $A^{ab}_1$ as
\begin{equation}
A_1^{ab}=-\frac{1}{80}\left(2e^{-2\phi_1}+2\kappa e^{-2\phi_2}+\lambda e^{4(\phi_1+\phi_2)}\right)\Omega^{ab}=-\frac{\mathcal{W}}{2}\Omega^{ab}.
\end{equation}
In the last equation \eqref{last_raw_BPS}, we have also used the composite connection 
\begin{equation}\label{SO(2)xSO(2)_Qz}
{Q_{z b}}^a=\frac{g}{2}\left[\mathbb{A}_1{(\Gamma_{12})_b}^a+\kappa\mathbb{A}_2{(\Gamma_{34})_b}^a\right].
\end{equation}
\indent To proceed, we impose the following projection conditions
\begin{equation}
{(\Gamma_{12})^a}_b\epsilon^b={(\Gamma_{34})^a}_b\epsilon^b=-i\epsilon^a\, .\label{SO(2)xSO(2)_Proj_con}
\end{equation}
We also note that, if one of the gauge fields vanishes, one of the projectors in \eqref{SO(2)xSO(2)_Proj_con} is not needed. Therefore, in this case, supersymmetry is enhanced as also pointed out in \cite{M5_disk_Couzen1}. Using the two projectors in \eqref{SO(2)xSO(2)_Proj_con} and defining
\begin{equation}
\mathcal{H}\Slash\,=\frac{\left(e^{2\phi_1}\mathbb{A}'_1+e^{2\phi_2}\mathbb{A}'_2\right)}{5\sqrt{h_1h_2}},
\end{equation}
we can rewrite the above BPS equations as
\begin{eqnarray}
0&=&2\partial_m\epsilon^a-\frac{1}{\sqrt{f}}\gamma_m\gamma^{\hat{\rho}}\epsilon^a+\frac{f'}{2f\sqrt{h_1}}\gamma_m\gamma^{\hat{r}}\epsilon^a-g\mathcal{W}\gamma_m\epsilon^a+2i\mathcal{H}\Slash\,\gamma_m\gamma^{\hat{r}\hat{z}}\epsilon^a,\label{BPS_m}\\
0&=&2\partial_\rho\epsilon^a+\frac{f'}{2f\sqrt{h_1}}\gamma_\rho\gamma^{\hat{r}}\epsilon^a-g\mathcal{W}\gamma_\rho\epsilon^a+2i\mathcal{H}\Slash\,\gamma_\rho\gamma^{\hat{r}\hat{z}}\epsilon^a,\label{BPS_rho}\\
0&=&2\partial_r\epsilon^a-g\mathcal{W}\gamma_r\epsilon^a-8i\sqrt{h_1}\mathcal{H}\Slash\,\gamma^{\hat{z}}\epsilon^a,\label{BPS_r}\\
0&=&2\partial_z\epsilon^a+ig\left(\mathbb{A}_1+\kappa \mathbb{A}_2\right)\epsilon^a+\frac{h'_2}{2h_2\sqrt{h_1}}\gamma_z\gamma^{\hat{r}}\epsilon^a-g\mathcal{W}\gamma_z\epsilon^a+8i\sqrt{h_2}\mathcal{H}\Slash\,\gamma^{\hat{r}}\epsilon^a\, .\ \ \qquad\label{BPS_z}
\end{eqnarray}
Similarly, the condition $\delta\chi^{abc}=0$ yields
\begin{eqnarray}
0&=&\phi'_i\gamma^r\epsilon^a-5g\,\mathbb{G}_{ij}\frac{\partial\mathcal{W}}{\partial\phi_j}\epsilon^a-10i\,\mathbb{G}_{ij}\frac{\partial\mathcal{\mathcal{H}\Slash\,}}{\partial\phi_j}\gamma^{\hat{r}\hat{z}}\epsilon^a\, .
\end{eqnarray}
\indent With the supersymmetry parameter \eqref{7DKilling}, the first two BPS equations given in \eqref{BPS_m} and \eqref{BPS_r} reduce to a single equation 
\begin{eqnarray}
0&=&\frac{2n_a}{\sqrt{f}}\left(\nabla_{\hat{\alpha}}^{AdS_5}\vartheta\right)\otimes\eta+\frac{f'}{2f\sqrt{h_1}}n_a\left(\mathbf{g}_{\hat{\alpha}}\vartheta\right)\otimes\left(i\sigma_2\eta\right)\nonumber\\&&-g\mathcal{W}n_a\left(\mathbf{g}_{\hat{\alpha}}\vartheta\right)\otimes\left(\sigma_3\eta\right)-2\mathcal{H}\Slash\,n_a\left(\mathbf{g}_{\hat{\alpha}}\vartheta\right)\otimes\eta\nonumber\\
&=&n_a\left(\mathbf{g}_{\hat{\alpha}}\vartheta\right)\otimes\left\{\frac{s}{\sqrt{f}}\eta+\frac{f'}{2f\sqrt{h_1}}(i\sigma_2)\eta-g\mathcal{W}\sigma_3\eta-2\mathcal{H}\Slash\,\eta\right\}\, .
\end{eqnarray}
We have used \eqref{5DKilling} and expressed all the supersymmetry parameters in terms of two- and four-component spinors. 
\\
\indent Repeating the same procedure for the remaining conditions, we find the following set of BPS equations on the two-component spinor $\eta$
\begin{eqnarray}
0&=&\frac{s}{\sqrt{f}}\eta+\frac{f'}{2f\sqrt{h_1}}(i\sigma_2)\eta-g\mathcal{W}\sigma_3\eta-2\mathcal{H}\Slash\,\eta,\label{AdS5_con}\\
0&=&\frac{2}{\sqrt{h_1}}\partial_r\eta-g\mathcal{W}\sigma_1\eta-8\mathcal{H}\Slash\,(i\sigma_2)\eta,\label{partial_r_con}\\
0&=&\frac{g\widehat{\mathbb{A}}}{\sqrt{h_2}}\eta-\frac{h'_2}{2h_2\sqrt{h_1}}\sigma_3\eta+g\mathcal{W}(i\sigma_2)\eta+8\mathcal{H}\Slash\,\sigma_1\eta,\label{partial_z_con}\\
0&=&\frac{\phi'_i}{\sqrt{h_1}}\sigma_1\eta-5g\,\mathbb{G}_{ij}\frac{\partial\mathcal{W}}{\partial\phi_j}\eta+10\,\mathbb{G}_{ij}\frac{\partial\mathcal{\mathcal{H}\Slash\,}}{\partial\phi_j}\sigma_3\eta\, .\label{phii_con}
\end{eqnarray}
In \eqref{partial_z_con}, we have defined 
\begin{equation}\label{AandA}
g\widehat{\mathbb{A}}=2q+g(\mathbb{A}_1+\kappa \mathbb{A}_2)
\end{equation}
with a constant $q$. As in \cite{Bah_M5}, this term is invariant under the transformations
\begin{equation}
\eta\rightarrow e^{i\alpha_0z}\eta,\qquad \mathbb{A}_1\rightarrow \mathbb{A}_1-\frac{\alpha_0}{g},\qquad \mathbb{A}_2\rightarrow \mathbb{A}_2-\frac{\kappa\alpha_0}{g}
\end{equation}
where $\alpha_0$ is an arbitrary constant. We then take the ansatz for the two-component spinor $\eta$ to be
\begin{equation}\label{eta_ansatz}
\eta(r,z)=e^{iqz}\widehat{\eta}(r)
\end{equation}
\indent The BPS equation given in \eqref{partial_r_con} determines the $\eta(r,z)$ as a function of $r$ while the remaining ones take the form $M^{(x)}\eta=0$ where $x=1,2,3,4$ respectively labels the BPS equations \eqref{AdS5_con}, \eqref{partial_z_con}, and \eqref{phii_con} with $i=1$ and $2$. The four $2\times2$ matrices $M^{(x)}$ can be parametrized by
\begin{equation}
M^{(x)}=X^{(x)}_0\mathds{1}_2+X^{(x)}_1\sigma^1+X^{(x)}_2(i\sigma^2)+X^{(x)}_3\sigma^3\, .
\end{equation}
Following \cite{Bah_M5}, for each matrix $M^{(x)}$, we define the two-component vectors
\begin{equation}
v^{(x)}=\begin{pmatrix} X^{(x)}_1+X^{(x)}_2 \\ -X^{(x)}_0-X^{(x)}_3 \end{pmatrix},\qquad w^{(x)}=\begin{pmatrix} X^{(x)}_0-X^{(x)}_3 \\ -X^{(x)}_1+X^{(x)}_2 \end{pmatrix}
\end{equation}
together with
\begin{equation}
\mathcal{A}^{xy}=\text{det}(v^{(x)}|w^{(y)}),\qquad\mathcal{B}^{xy}=\text{det}(v^{(x)}|v^{(y)}),\qquad\mathcal{C}^{xy}=\text{det}(w^{(x)}|w^{(y)}).
\end{equation}
The notation $(a|b)$ denotes a $2\times2$ matrix constructed from the two-column vectors $a$ and $b$. We can find necessary conditions for the existence of a non-trivial solution for $\eta$ by setting $\mathcal{A}^{xy}$, $\mathcal{B}^{xy}$, and $\mathcal{C}^{xy}$ to zero.
\\
\indent Starting from the first matrix $\mathcal{A}^{xy}$, the vanishing of the diagonal components $\mathcal{A}^{xx}$ gives the following conditions
\begin{eqnarray}
0&=&\frac{4s\mathcal{H}\Slash\,}{\sqrt{f}}+g^2\mathcal{W}^2-\frac{1}{f}-\frac{(f')^2}{4h_1f^2}-4\mathcal{H}\Slash\,^2,\label{DABPS1}\\
0&=&g^2\widehat{\mathbb{A}}^2-\frac{(h'_2)^2}{4h_1h_2}-(64\mathcal{H}\Slash\,^2-g^2\mathcal{W}^2)h_2,\label{DABPS2}\\
0&=&\frac{(\phi'_i)^2}{25h_1}-g^2\mathcal{W}_i{}^2+4\mathcal{H}\Slash\,_i{}^2\label{DABPS3}
\end{eqnarray}
in which, for simplicity, we have written
\begin{equation}
\mathcal{W}_i=\mathbb{G}_{ij}\frac{\partial\mathcal{W}}{\partial\phi_j}\qquad\text{ and }\qquad\mathcal{H}\Slash\,_i=\mathbb{G}_{ij}\frac{\partial\mathcal{\mathcal{H}\Slash\,}}{\partial\phi_j}\, .
\end{equation}
Note that \eqref{DABPS3} indeed contains two different conditions for $i=1$ and $2$. In the following, we also write doublets of conditions, such as those given in \eqref{DABPS3}, with a subscript $i$ for convenience. 
\\
\indent The vanishing of the off-diagonal symmetric components $\mathcal{A}^{xy}+\mathcal{A}^{yx}$ ($x\neq y$) yields 
\begin{eqnarray}
0&=&2\left(\frac{s}{\sqrt{f}}-2\mathcal{H}\Slash\,\right)\frac{g\widehat{\mathbb{A}}}{\sqrt{h_2}}+\left(\frac{f'}{f}-\frac{h'_2}{h_2}\right)\frac{g\mathcal{W}}{\sqrt{h_1}},\label{SABPS1}\\
0&=&g\mathcal{W}_i\left(\frac{s}{\sqrt{f}}-2\mathcal{H}\Slash\,\right)-2g\mathcal{W}\mathcal{H}\Slash\,_i,\label{SABPS2}\\
0&=&\frac{5g^2\mathcal{W}_i\widehat{\mathbb{A}}}{\sqrt{h_2}}-\frac{1}{\sqrt{h_1}}\left(\frac{5h'_2\mathcal{H}\Slash\,_i}{h_2}-8\phi'_i\mathcal{H}\Slash\,\right),\qquad\label{SABPS3}\\
0&=&\frac{\phi'_1\phi'_2}{25h_1}-g^2\mathcal{W}_1\mathcal{W}_2+4\mathcal{H}\Slash\,_1\mathcal{H}\Slash\,_2\qquad\label{SABPS4}
\end{eqnarray}
while the off-diagonal antisymmetric components $\mathcal{A}^{xy}-\mathcal{A}^{yx}=0$ give
\begin{eqnarray}
0&=&\frac{2g^2\mathcal{W}\widehat{\mathbb{A}}}{\sqrt{h_2}}+\frac{8f'\mathcal{H}\Slash\,}{f\sqrt{h_1}}-\left(\frac{s}{\sqrt{f}}-2\mathcal{H}\Slash\,\right)\frac{h'_2}{h_2\sqrt{h_1}},\label{AABPS1}\\
0&=&20\left(\frac{s}{\sqrt{f}}-2\mathcal{H}\Slash\,\right)\mathcal{H}\Slash\,_i+\frac{f'\phi'_i}{fh_1}-10g^2\mathcal{W}_i\mathcal{W},\label{AABPS2}\\
0&=&\frac{20g\widehat{\mathbb{A}}\mathcal{H}\Slash\,_i}{\sqrt{h_2}}-\frac{5gh'_2\mathcal{W}_i}{h_2\sqrt{h_1}}+\frac{2g\phi'_i\mathcal{W}}{\sqrt{h_1}},\label{AABPS3}\\
0&=&g\left(\mathcal{H}\Slash\,_1\mathcal{W}_2-\mathcal{H}\Slash\,_2\mathcal{W}_1\right).\label{AABPS4}
\end{eqnarray}
\indent Matrices $\mathcal{B}^{xy}$ and $\mathcal{C}^{xy}$ are both antisymmetric in $xy$ indices, and it turns out that the resulting BPS conditions arising from the combinations $\mathcal{B}^{xy}\pm\mathcal{C}^{xy}=0$ take a simpler form. The combinations $\mathcal{B}^{xy}+\mathcal{C}^{xy}=0$ give the following conditions
\begin{eqnarray}
0&=&20\mathcal{W}\mathcal{H}\Slash\,-\frac{2s\mathcal{W}}{\sqrt{f}}+\frac{f'\widehat{\mathbb{A}}}{f\sqrt{h_1h_2}},\label{ABCBPS1}\\
0&=&2\phi'_if\mathcal{W}-5f'\mathcal{W}_i,\label{ABCBPS2}\\
0&=&16\mathcal{H}\Slash\,_i\mathcal{H}\Slash\,+\frac{h'_2\phi'_i}{10h_1h_2}-g^2\mathcal{W}_i\mathcal{W},\label{ABCBPS3}\\
0&=&\phi'_1\mathcal{H}\Slash\,_2-\phi'_2\mathcal{H}\Slash\,_1\label{ABCBPS4}
\end{eqnarray}
while the combinations $\mathcal{B}^{xy}-\mathcal{C}^{xy}=0$ lead to
\begin{eqnarray}
0&=&32\left(\frac{s}{\sqrt{f}}-2\mathcal{H}\Slash\,\right)\mathcal{H}\Slash\,+\frac{f'h'_2}{fh_1h_2}-4g^2\mathcal{W}^2,\label{SBCBPS1}\\
0&=&\left(\frac{s}{\sqrt{f}}-2\mathcal{H}\Slash\,\right)\phi'_i-\frac{5f'\mathcal{H}\Slash\,_i}{f},\label{SBCBPS2}\\
0&=&\mathcal{H}\Slash\,_i\mathcal{W}-4\mathcal{W}_i\mathcal{H}\Slash\,-\frac{\phi'_i\widehat{\mathbb{A}}}{10\sqrt{h_1h_2}},\label{SBCBPS3}\\
0&=&\phi'_1\mathcal{W}_2-\phi'_2\mathcal{W}_1\, .\label{SBCBPS4}
\end{eqnarray}
There are, in total, 28 algebraic conditions obtained from the BPS equations \eqref{AdS5_con}, \eqref{partial_z_con}, and \eqref{phii_con}.  
\\
\indent Substituting the explicit forms of  $\mathcal{W}_i$ and $\mathcal{H}\Slash\,_i$ together with $\mathbb{A}'_1$ and $\mathbb{A}'_2$ from \eqref{Gen_SO(2)xSO(2)_Ansatz} into \eqref{AABPS4}, we find 
\begin{equation}\label{1st_Strongest_con}
a_1 \left(\lambda  e^{4 \phi_1+6\phi_2}-\kappa \right)+a_2 \left(1-\lambda  e^{6\phi_1+4\phi_2}\right)=0\, .
\end{equation}
This can be readily solved for $\phi_2$ in terms of $\phi_1$ as follows 
\begin{equation}
\phi_2=-2\phi_1+\frac{1}{2}\ln\left[\frac{a_2 e^{6\phi_1} \left(2^{\frac{1}{3}} a_2 \lambda  e^{6 \phi_1}+Z\right)}{3a_1 Z}+\frac{2^{\frac{2}{3}} Z}{6a_1\lambda }\right]\label{Gen_phi2_soln}
\end{equation}
where
\begin{eqnarray}
Z&=&\left[3 \sqrt{3} \sqrt{a_1^2 \lambda ^4 \left(a_1 \kappa -a_2\right) e^{16 \phi_1} \left[27 a_1^2 \left(a_1 \kappa -a_2\right)+4 a_2^3 \lambda  e^{10 \phi_1}\right]}\right.\nonumber\\
&&\left.\phantom{\frac{1}{2}}+27 a_1^2 \lambda ^2 \left(a_1 \kappa -a_2\right) e^{8 \phi_1}+2 a_2^3 \lambda ^3 e^{18 \phi_1}\right]^{\frac{1}{3}}\, .\label{Z_def}
\end{eqnarray}
In addition, we can find the solutions for $f$ and $h_1$ from \eqref{SABPS2} and \eqref{DABPS3} for $i=1$
\begin{eqnarray}
f&=&\sqrt{\frac{a_1 \left(2 \kappa +\lambda  e^{4 \phi_1+6 \phi_2}\right)-2 a_2}{s\left(2 \kappa  e^{2 \phi_1}+\lambda  e^{6 (\phi_1+\phi_2)}-3 e^{2 \phi_2}\right)}},\label{f_gen_sol}\\
h_1&=&\frac{400 f^5 e^{4 (\phi_1+\phi_2)} (\phi'_1)^2}{g^2 f^5 \left(2 \kappa  e^{2 \phi_1}+\lambda  e^{6 (\phi_1+\phi_2)}-3 e^{2\phi_2}\right)^2-64 \left(3 a_1 e^{2 \phi_2}-2a_2 e^{2 \phi_1}\right)^2}\, .\qquad\quad\ \label{h1_gen_sol}
\end{eqnarray}
\indent We now determine the function $\widehat{\mathbb{A}}$ defined in \eqref{AandA}. From \eqref{SBCBPS3} with $i=1$, we find the following expression for $\sqrt{h_1h_2}$
\begin{equation}
\small
\sqrt{h_1h_2}=\frac{20 f^{5/2} \widehat{\mathbb{A}} e^{4 (\phi_1+\phi_2)} \phi'_1}{a_1 \left(2 \kappa  e^{2 (\phi_1+\phi_2)}+\lambda  e^{6 \phi_1+8 \phi_2}-6 e^{4 \phi_2}\right)+2 a_2 \left(2 \kappa  e^{4 \phi_1}+\lambda  e^{8\phi_1+6\phi_2}-2 e^{2(\phi_1+\phi_2)}\right)}\, .\label{sqrth1h1_1}
\end{equation}
With $\widehat{\mathbb{A}}'=\mathbb{A}'_1+\kappa\mathbb{A}'_2$, we also find another expression for $\sqrt{h_1h_2}$ from \eqref{Gen_SO(2)xSO(2)_Ansatz} as
\begin{equation}
\sqrt{h_1h_2}=\frac{f^{5/2} \widehat{\mathbb{A}}' e^{4 (\phi_1+\phi_2)}}{a_2 \kappa  e^{4\phi_1}+a_1 e^{4 \phi_2}}.\label{sqrth1h1_2}
\end{equation}
Equating \eqref{sqrth1h1_1} and \eqref{sqrth1h1_2}, we find an ordinary differential equation
\begin{equation}\small
\widehat{\mathbb{A}}'=\frac{20 \widehat{\mathbb{A}}\phi'_1 \left(a_2 \kappa  e^{4 \phi_1}+a_1 e^{4 \phi_2}\right)}{a_1 \left(2 \kappa  e^{2 (\phi_1+\phi_2)}+\lambda  e^{6 \phi_1+8\phi_2}-6 e^{4\phi_2}\right)+2 a_2 \left(2 \kappa  e^{4 \phi_1}+\lambda  e^{8\phi_1+6\phi_2}-2 e^{2(\phi_1+\phi_2)}\right)}\, .\label{gen_A_prime}
\end{equation}
With the $\phi_2$ solution in \eqref{Gen_phi2_soln}, this equation can give a solution for $\widehat{\mathbb{A}}$ as a function of $\phi_1$. However, we are not able to analytically integrate this equation to find $\widehat{\mathbb{A}}(\phi_1)$ in full generality. Although a numerical analysis can lead to an explicit solution, we instead impose some extra condition in such a way that equation \eqref{gen_A_prime} can be analytically integrated. The resulting solution could be more useful than the numerical one.  

\subsection{Regular solutions}
We now consider a simplified $SO(2)\times SO(2)$ symmetric solution obtained by imposing the condition 
\begin{equation}
a_2=-\kappa a_1\, . 
\end{equation}
This is similar to the analysis of \cite{7D_N2_disk} for the $AdS_5\times \Sigma$ solutions in half-maximal gauged supergravity. For convenience, we will also redefine the scalar fields as
\begin{equation}
\sigma=2(\phi_1+\phi_2)\qquad\text{ and }\qquad\phi=\phi_1-\phi_2\, .
\end{equation}
With all these, the condition \eqref{1st_Strongest_con} becomes
\begin{equation}
a_1 e^{-\sigma-\phi} \left(\kappa  \lambda  e^{\frac{5 \sigma}{2}+2 \phi}-2 \kappa  e^{\phi}+\lambda  e^{\frac{5\sigma}{2}}\right)=0\, .\label{SO(2)xSO(2)_Strongest_con}
\end{equation}
This can be solved for a real solution of $\phi$ in terms of $\sigma$ or vice versa only in $SO(5)$ and $SO(3,2)$ gauge groups with $\kappa=\lambda=\pm1$. For $\kappa=\lambda$, the solution for $\phi$ is then given by
\begin{equation}
\phi=\ln \left[\lambda  e^{-\frac{5}{2}\sigma}+\sqrt{e^{-5 \sigma}-\lambda }\right].
\end{equation}
Note that, in the $SO(5)$ gauge group with $\lambda=1$, reality of the $\phi$ solution requires $\sigma<0$.
\\
\indent With the resulting solution for $\phi$, we can determine $f$ and $h_1$ from \eqref{f_gen_sol} and \eqref{h1_gen_sol} as 
\begin{eqnarray}
f&=&\frac{64B e^{\sigma}}{g^2 \left(1-\lambda e^{5\sigma}\right)^{\frac{1}{4}}},\label{f_SO(2)xSO(2)_sol}\\
h_1&=&\frac{25B e^{6 \sigma} \left(\sigma'\right)^2}{g^2\left(1-\lambda e^{5 \sigma} \right)^2 \left[B-e^{-5 \sigma }\left(1-\lambda  e^{5 \sigma}\right)^{\frac{1}{4}}\right]}\label{h1_SO(2)xSO(2)_sol}
\end{eqnarray}
with $B=\frac{g^2\sqrt{a_1}}{64\sqrt{s}}$. The differential equation \eqref{gen_A_prime} now takes a much simpler form
\begin{equation}
\widehat{\mathbb{A}}'=\frac{20 \widehat{\mathbb{A}} \sigma'}{3 \lambda  e^{5 \sigma}-4}\label{SO(2)xSO(2)_A_prime}
\end{equation}
which can be solved by
\begin{equation}
\widehat{\mathbb{A}}=c_1 \left(3 -4 \lambda e^{-5 \sigma}\right)
\end{equation}
for an integration constant $c_1$. 
\\
\indent With all these, the solution for $h_2$ can be obtained from \eqref{sqrth1h1_1} as
\begin{equation}
h_2=\frac{64c_1{}^2 e^{\sigma}\left[B-e^{-5 \sigma }\left(1-\lambda  e^{5 \sigma}\right)^{\frac{1}{4}}\right]}{\left(1-\lambda  e^{5 \sigma}\right)^{\frac{1}{4}}}\, .
\end{equation}
By solving equation \eqref{partial_r_con}, we find that the explicit form of the Killing spinor is given by
\begin{equation}
\eta=\frac{Y_0 e^{i qz}e^{\frac{\sigma}{4}} }{\left(1-\lambda e^{5 \sigma}\right)^{\frac{1}{16}}}\begin{pmatrix} \sqrt{\sqrt{B}+s e^{-\frac{5\sigma}{2}} \left(1-\lambda  e^{5 \sigma}\right)^{\frac{1}{8}}} \\ -\sqrt{\sqrt{B}-s e^{-\frac{5\sigma}{2}} \left(1-\lambda  e^{5 \sigma}\right)^{\frac{1}{8}}}\end{pmatrix}\label{SO(2)xSO(2)_eta}
\end{equation}
with $Y_0$ being an integration constant.
\\
\indent In order to find an explicit form of the $SO(2)\times SO(2)$ gauge fields, we substitute the above solution into the differential equations \eqref{Gen_SO(2)xSO(2)_Ansatz} and find
\begin{eqnarray}
\mathbb{A}_1&=&\frac{c_1}{2}  \left[3-4 e^{-5 \sigma}\left(\lambda  - e^{\frac{5 \sigma}{2}} \sqrt{e^{-5 \sigma}-\lambda }\right)\right]-\frac{q}{g}+\frac{c_2}{4},\\ 
\mathbb{A}_2&=&\frac{c_1 }{2} \left[3 \lambda -4 e^{-5 \sigma} \left(\lambda  e^{\frac{5 \sigma}{2}} \sqrt{e^{-5 \sigma}-\lambda }+1\right)\right]-\frac{\lambda q}{g}-\frac{\lambda c_2 }{4}
\end{eqnarray}
where $c_2$ is an integration constant. 
\\
\indent With the following identification
\begin{equation}
g=32h,\qquad \lambda=\pm m^2=\pm\frac{\widetilde{g}_1{}^2}{(16h)^2}=\pm\frac{(16h)^2}{\widetilde{g}_1{}^2},\qquad c_1=\frac{\mathcal{C}}{16h}\label{SO(2)xSO(2)_map}
\end{equation}
for $m$, $\widetilde{g}_1$, $h$, and $\mathcal{C}$ being the parameters used in \cite{7D_N2_disk}, it can be verified that the solution takes the same form as $SO(2)\times SO(2)$ symmetric solution found in \cite{7D_N2_disk} from $N=2$ gauged supergravity coupled to vector multiplets. In particular, the solution from $SO(5)$ ($\lambda=1$) gauge group is equivalent to the solution from $SO(4)$ gauge group studied in \cite{7D_N2_disk} with $\widetilde{g}_1{}^2=(16h)^2$ and 
\begin{equation}\label{SO(2)xSO(2)_map_to_R}
A_1=2(\mathbb{A}_1+\mathbb{A}_2)\qquad\text{ and }\qquad A_2=2(\mathbb{A}_1-\mathbb{A}_2)
\end{equation}
in which $A_1$ and $A_2$ are the $SO(2)\times SO(2)$ gauge fields in \cite{7D_N2_disk}. On the other hands, the solution in $SO(3,2)$ ($\lambda=-1$) gauge group is equivalent the $SO(2)\times SO(2)$ symmetric solution in $SO(2,2)$ gauge group considered in \cite{7D_N2_disk} with $\widetilde{g}_1{}^2=(16h)^2$ and 
\begin{equation}
A_1=2(\mathbb{A}_1-\mathbb{A}_2)\qquad\text{ and }\qquad A_2=2(\mathbb{A}_1+\mathbb{A}_2).
\end{equation}
As pointed out in \cite{7D_N2_disk}, the two gauge fields $A_1$ and $A_2$ are interchanged in $SO(2,2)$ gauge group. However, in half-maximal gauged supergravity, these solutions currently have no known higher-dimensional origin except for particular values of the two gauge coupling constants in $SO(4)\sim SO(3)\times SO(3)$ gauge group to be equal. The uplifted solution even in the case of $SO(4)$ gauge group has not been considered in \cite{7D_N2_disk}. Therefore, in the present paper, we will close this gap by uplifting the solution to eleven-dimensional supergravity. 
\\
\indent Since the solutions take the same form as those considered in \cite{7D_N2_disk}, we can proceed to determine the range of the coordinate $r$ that leads to regular solutions as in \cite{7D_N2_disk}. The analysis is essentially the same as that performed in \cite{7D_N2_disk}. Accordingly, we will only repeat the main results and refer to \cite{7D_N2_disk} for more detail. We first fix the form of the undetermined $\sigma$ solution to be
\begin{equation}\label{1/5sigma_soln}
\sigma=-\frac{1}{5}\ln r
\end{equation}
for $r>0$. The seven-dimensional metric then reads
\begin{equation}
ds^2_7=\frac{64 B r^{\frac{1}{20}}}{g^2 \left(r-\lambda\right)^{\frac{1}{4}}}\left[ds^2_{AdS_5}+\frac{r^{-\frac{5}{4}}}{64W\left(r-\lambda\right)^{\frac{7}{4}}}dr^2+\frac{g^2c_1{}^2W}{B}dz^2\right]\label{SO(2,2)_SO(2)xSO(2)_Met}
\end{equation}
with
\begin{equation}\label{SO(2)xSO(2)_Def_W}
W=B-r^{\frac{3}{4}}\left(r-\lambda\right)^{\frac{1}{4}}
\end{equation}
while the solutions for $\phi$, $\mathbb{A}_1$, and $\mathbb{A}_2$ are given by
\begin{eqnarray}
\phi&=&\ln \left[\lambda  \sqrt{r}+\sqrt{r-\lambda }\right],\\
\mathbb{A}_1&=&\frac{c_1}{2}\left(3-4 \lambda  r+4 \sqrt{r} \sqrt{r-\lambda }\right)-\frac{q}{g}+\frac{c_2}{4} ,\\ 
\mathbb{A}_2&=&\frac{c_1}{2} \left(3 \lambda -4 r-4 \lambda  \sqrt{r} \sqrt{r-\lambda }\right)-\frac{\lambda  q}{g}-\frac{\lambda  c_2}{4}\, .
\end{eqnarray}
The Killing spinor $\eta$ in terms of $r$ takes the form
\begin{equation}\label{SO(2)xSO(2)_eta_R}
\eta=Y_0e^{iqz}\frac{r^{\frac{1}{80}}}{(r-\lambda)^{\frac{1}{16}}}\begin{pmatrix} \sqrt{\sqrt{B}+sr^{\frac{3}{8}}(r-\lambda)^{\frac{1}{8}}} \\ -\sqrt{\sqrt{B}-sr^{\frac{3}{8}}(r-\lambda)^{\frac{1}{8}}} \end{pmatrix}.
\end{equation}

\subsubsection{Regular solutions from $SO(5)$ gauge group}\label{Y_SO(2)xSO(2)_SO(5)_Sec}
\begin{figure}[h!]
  \centering
  \begin{subfigure}[b]{0.326\linewidth}
    \includegraphics[width=\linewidth]{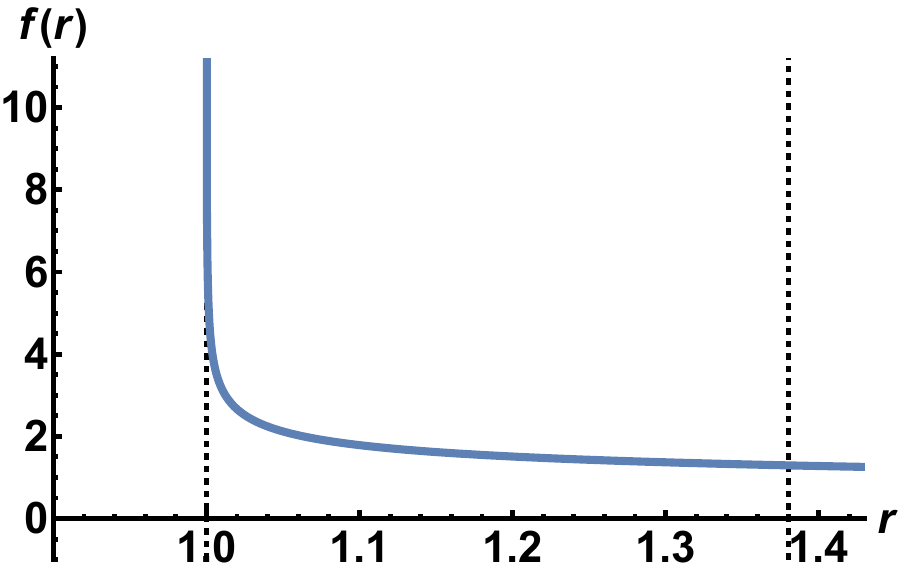}
\caption{$f$ solution}
  \end{subfigure}
  \begin{subfigure}[b]{0.326\linewidth}
    \includegraphics[width=\linewidth]{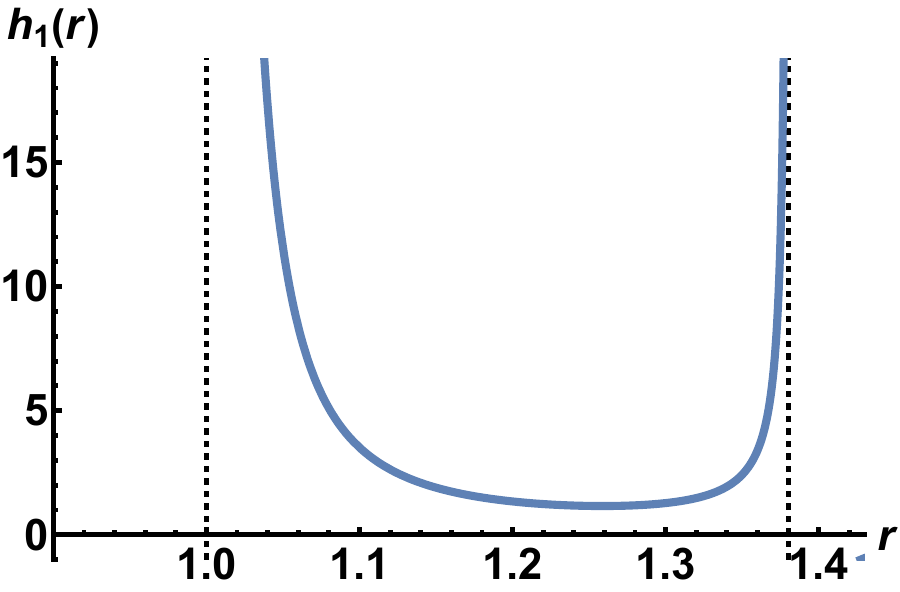}
\caption{$h_1$ solution}
  \end{subfigure}
  \begin{subfigure}[b]{0.326\linewidth}
    \includegraphics[width=\linewidth]{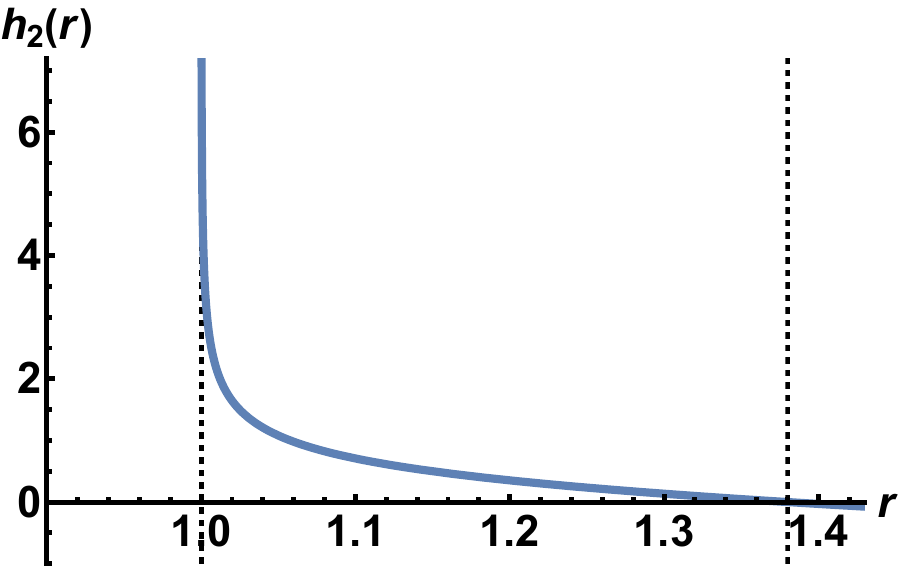}
\caption{$h_2$ solution}
  \end{subfigure}
  \caption{Numerical plots of the warp factors for $SO(2)\times SO(2)$ symmetric solution in $SO(5)$ gauge group with $B=1$, $c_1=\frac{1}{8}$, and $g=8$. The solution is regular in the range $1<r<r_1=1.38$ with the two vertical dashed lines representing the two boundaries.}
  \label{Fig14}
\end{figure}

For $\lambda=1$, there is only one possible range for the radial coordinate $r$ in order to obtain a regular solution. This is given by 
\begin{equation}\label{SO(5)_SO(2)xSO(2)_Pos_Range}
1<r<r_1
\end{equation}
with $B>0$. $r_1$ is determined from $W(r_1)=0$ as
\begin{equation}\label{SO(2)xSO(2)_r_roots}
r_1=\frac{1}{4}+\frac{\sqrt{X}}{2}+\frac{1}{2} \sqrt{\frac{3}{4}-X+\frac{1}{4 \sqrt{X}}}
\end{equation}
for
\begin{equation}
X=\frac{1}{4}-\frac{2^{\frac{7}{3}} B^4}{3^{\frac{1}{3}}\left(\sqrt{768 B^{12}+81 B^8}-9 B^4\right)^{\frac{1}{3}}}+\frac{\left(\sqrt{768 B^{12}+81 B^8}-9 B^4\right)^{\frac{1}{3}}}{2^{\frac{1}{3}} 3^{\frac{2}{3}}}\, .\label{X_def1}
\end{equation}
An example of numerical plots for the three warp factors is shown in figure \ref{Fig14}.

\begin{figure}[h!]
  \centering
    \includegraphics[width=0.5\linewidth]{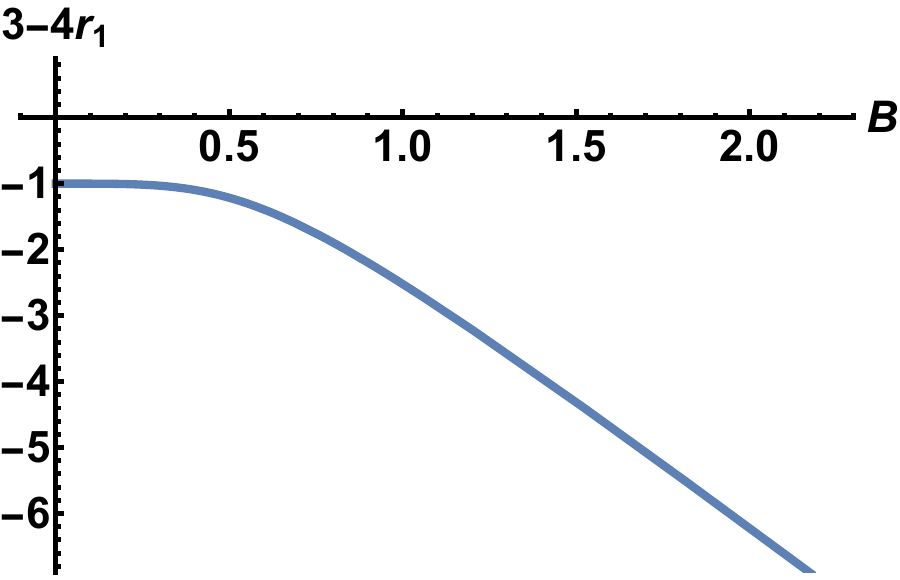}
  \caption{A numerical plot of the function $3-4r_1$ appearing in the condition \eqref{SO(5)_SO(2)xSO(2)_smooth_con}.}
  \label{Fig15}
\end{figure}

As $r\rightarrow 1$, the seven-dimensional metric is conformal to a product of $AdS_5$ and a cylinder 
\begin{equation}\label{SO(2)d_rto1_metric}
ds_7^2\approx\frac{64}{c_1{}^2 g^4 R^2}\left[ds^2_{AdS_5}+c_1{}^2 g^2(dR^2+dz^2)\right]
\end{equation}
with the new radial coordinate $R$ defined by $r=1+B^4 c_1{}^8 g^8 R^8$. In this limit, the circle parametrized by the coordinate $z$ decompactifies, and the metric is asymptotic to $AdS_7$ space with $AdS_5\times S^1$ boundary. The decompactification of the $z$-circle leads to infinite holographic central charge for the $N=1$ SCFT dual to the $AdS_5$ factor. As pointed out in \cite{Gutperle_coD2_defect}, the solution should be holographically interpreted as a codimension-2 conformal defect within the $N=(2,0)$ six-dimensional SCFT dual to the $AdS_7$ vacuum rather than a compactification of the $N=(2,0)$ SCFT on $\Sigma$ to four dimensions.  
\\
\indent On the other hand, as $r\rightarrow r_1$, the seven-dimensional metric is approximately given by
\begin{equation}\label{SO(5)_SO(2)xSO(2)_spindle_metric}
ds_7^2\approx\frac{64B\,r_1^{\frac{1}{20}}}{g^2(r_1-1)^{\frac{1}{4}}}\left[ds^2_{AdS_5}+\frac{d\rho^2+c_1{}^2g^2\rho^2\left(3-4r_1\right)^2dz^2}{-4W'(r_1)r_1^{\frac{5}{4}}(r_1-1)^{\frac{7}{4}}}\right]
\end{equation}
with the coordinate $\rho$ defined by $\rho=\frac{1}{2}\sqrt{r_1-r}$. In this case, the $z$-circle shrinks as $\rho\rightarrow 0$. In order to make the $z$-circle shrink smoothly, we impose the condition
\begin{equation}\label{SO(5)_SO(2)xSO(2)_smooth_con}
c_1=\frac{-1}{gl\left(3-4r_1\right)},\qquad l=1,2,3,\ldots 
\end{equation}
giving rise to an $\mathbb{R}^2/\mathbb{Z}_l$ orbifold at $r=r_1$. Since the explicit form of $3-4r_1$ derived from \eqref{SO(2)xSO(2)_r_roots} is rather complicated, we only numerically show that this function is negative and less than $-1$ in the regularity range by a numerical plot in figure \ref{Fig15}. We have added $-1$ in the condition \eqref{SO(5)_SO(2)xSO(2)_smooth_con} to ensure positivity of $c_1$ for $g>0$. 

We can also compute the Euler Characteristic of $\Sigma$ by using the Gauss-Bonnet theorem and the fact that the boundary is geodesic. The result is given by
\begin{eqnarray}
\chi(\Sigma)&=&\frac{1}{4\pi}\int_{\Sigma}R_{\Sigma}\text{vol}_{\Sigma}\ =\ \frac{c_1 g \left(4 r_1-3\right)(r_1-1)^{\frac{1}{8}} r_1^{\frac{3}{8}}}{\sqrt{B}}\nonumber \\
&=&c_1 g \left(4 r_1-3\right)\ =\ \frac{1}{l}\label{SO(2)xSO(2)_SO(5)gg_chi}
\end{eqnarray}
as expected for a disk with a $Z_l$ orbifold singularity at the center. In order to make the $U(1)$ gauge fields on the boundary, $\mathbb{A}_1+\mathbb{A}_2$ and $\mathbb{A}_1-\mathbb{A}_2$, vanish identically at $r=r_1$, we choose the following choice of constants
\begin{equation}
q=\frac{gc_1}{2}\left(3-4r_1\right)=-\frac{1}{2l},\qquad c_2=-8 c_1 \sqrt{r_1(r_1-1)}\, .
\end{equation}
This leads to
\begin{eqnarray}
\mathbb{A}_1+\mathbb{A}_2&=&4c_1(r_1-r),\label{A121}\\ 
\mathbb{A}_1-\mathbb{A}_2&=&4 c_1 \left(\sqrt{r(r-1)}-\sqrt{r_1(r_1-1)}\right)\label{A122}
\end{eqnarray}
that vanish at $r=r_1$. 
\\
\indent Finally, the Killing spinor at the endpoint $r=r_1$ is explicitly given by
\begin{equation}\label{SO(32)_SO(2)xSO(2)_eta_Soln}
\eta= Y_0e^{-\frac{i z }{2 l}}r_1^{\frac{1}{5}} \begin{footnotesize}\begin{pmatrix} \sqrt{s+1} \\ -\sqrt{s-1} \end{pmatrix}\end{footnotesize}.
\end{equation}

\subsubsection{Regular solutions in $SO(3,2)$ gauge group}\label{Y_SO(2)xSO(2)_SO(3,2)_Sec} 
\begin{figure}[h!]
  \centering
  \begin{subfigure}[b]{0.326\linewidth}
    \includegraphics[width=\linewidth]{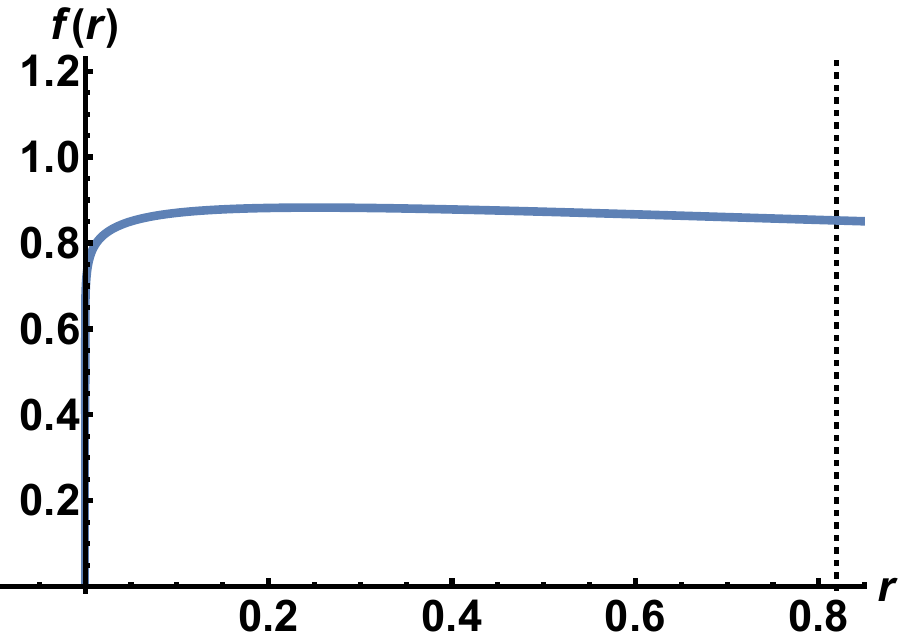}
\caption{$f$ solution}
  \end{subfigure}
  \begin{subfigure}[b]{0.326\linewidth}
    \includegraphics[width=\linewidth]{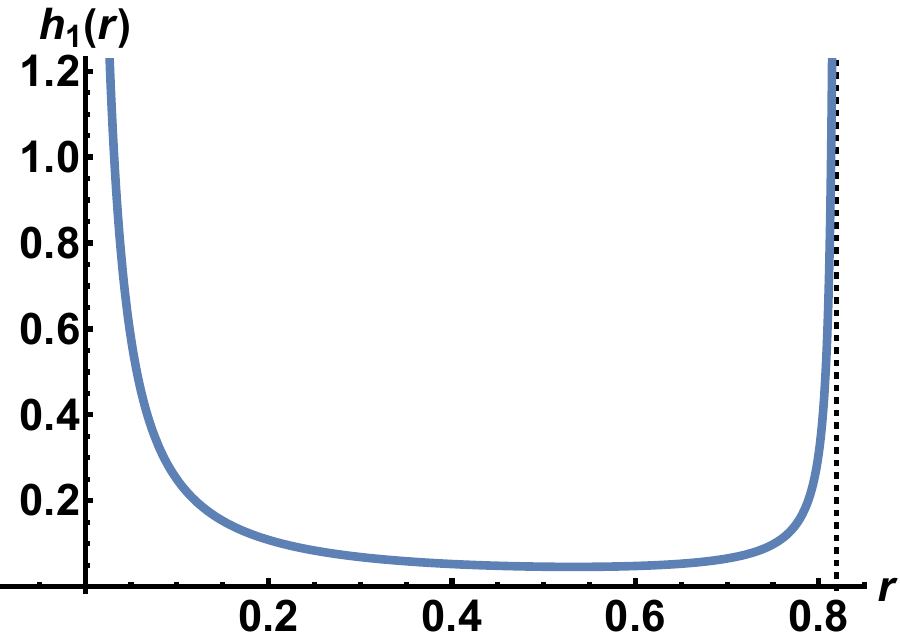}
\caption{$h_1$ solution}
  \end{subfigure}
  \begin{subfigure}[b]{0.326\linewidth}
    \includegraphics[width=\linewidth]{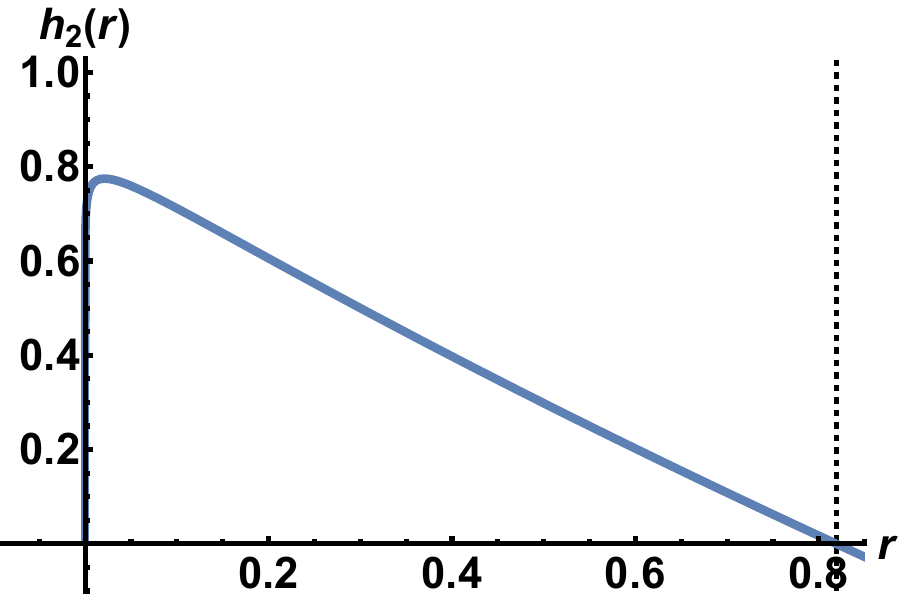}
\caption{$h_2$ solution}
  \end{subfigure}
  \caption{Numerical plots of the warp factors for $SO(2)\times SO(2)$ symmetric solution in $SO(3,2)$ gauge group with $B=1$, $c_1=\frac{1}{8}$, and $g=8$. The solution is regular in the range $0<r<r_2=0.819$ with the vertical dashed lines representing the boundary at $r=r_2=0.819$.}
  \label{Fig16}
\end{figure}

As in the previous case, for $SO(3,2)$ gauge group with $\lambda=-1$, there is only one possible range for the radial coordinate $r$ for a regular solution to exist 
\begin{equation}\label{SO(3,2)_SO(2)xSO(2)_Pos_Range}
 0<r<r_2
\end{equation}
for $B>0$. The endpoint $r_2$ is given by 
\begin{equation}\label{SO(2)xSO(2)_r2_roots}
r_2=-\frac{1}{4}-\frac{\sqrt{X}}{2}+\frac{1}{2} \sqrt{\frac{3}{4}-X+\frac{1}{4 \sqrt{X}}}
\end{equation}
with $X$ given in \eqref{X_def1}. An example of numerical plots for the three warp factors is shown in figure \ref{Fig16}.

\begin{figure}[h!]
  \centering
    \includegraphics[width=0.5\linewidth]{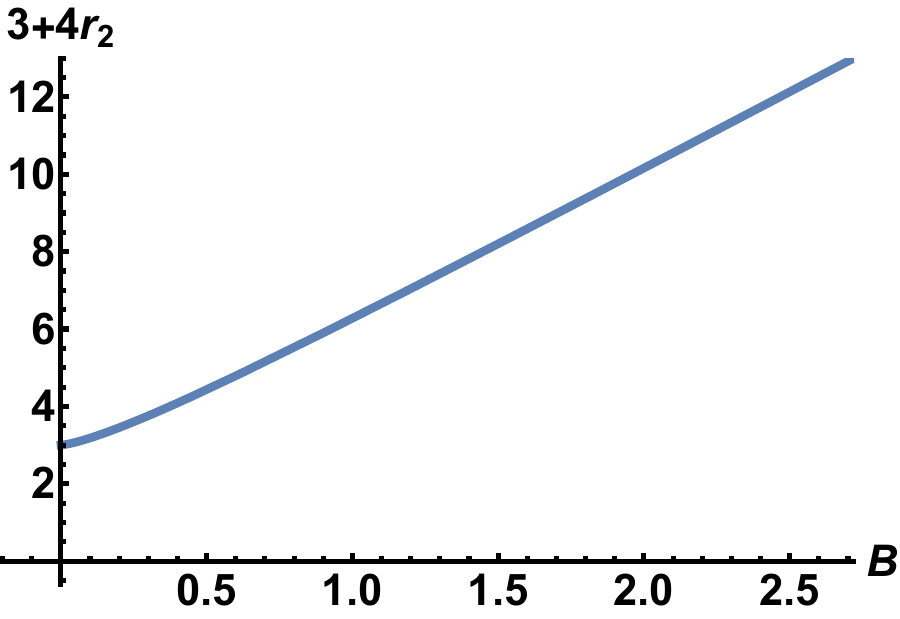}
  \caption{A numerical plot of the function $3+4r_2$ appearing in the condition \eqref{SO(32)_SO(2)xSO(2)_smooth_con}.}
  \label{Fig17}
\end{figure}

As $r\rightarrow 0$, the seven-dimensional metric is again conformal to a product of $AdS_5$ and a cylinder. With the new radial coordinate $R$ defined by $r=(3\sqrt{B}c_1{}g R) ^{\frac{8}{3}}$, the metric near $R=0$ becomes  
\begin{equation}\label{SO(2)d_rto0_metric}
ds_7^2\approx\frac{64(3B^8c_1 g R){}^{\frac{2}{15}}}{g^2}\left[ds^2_{AdS_5}+c_1{}^2 g^2(dR^2+dz^2)\right].
\end{equation}
As $r\rightarrow r_2$, the seven-dimensional metric is approximately given by
\begin{equation}\label{SO(32)_SO(2)xSO(2)_spindle_metric}
ds_7^2\approx\frac{64B\,r_2^{\frac{1}{20}}}{g^2(r_2+1)^{\frac{1}{4}}}\left[ds^2_{AdS_5}+\frac{d\rho^2+c_1{}^2g^2\rho^2\left(3+4r_{(-,+)}\right)^2dz^2}{-4W'(r_2)r_2^{\frac{5}{4}}(r_2+1)^{\frac{7}{4}}}\right]
\end{equation}
with the coordinate $\rho$ defined by $\rho=\frac{1}{2}\sqrt{r_2-r}$. The $z$ circle shrinks smoothly giving rise to an $\mathbb{R}^2/\mathbb{Z}_l$ orbifold at $r=r_2$ after imposing the condition
\begin{equation}\label{SO(32)_SO(2)xSO(2)_smooth_con}
c_1=\frac{1}{gl\left(3+4r_2\right)},\qquad l=1,2,3,\ldots \, .
\end{equation}
As in the previous case, the explicit form of $3+4r_2$ derived from \eqref{SO(2)xSO(2)_r2_roots} is rather complicated. Accordingly, we only numerically show that this function is always greater than $3$ in the regularity range by giving a numerical plot in figure \ref{Fig17}.

By choosing the constants to be
\begin{equation}
q=\frac{gc_1}{2}\left(3+4r_2\right)=\frac{1}{2l},\qquad c_2=-8 c_1 \sqrt{r_2(r_2+1)},
\end{equation} 
we find the $SO(2)\times SO(2)$ gauge fields
\begin{eqnarray}
\mathbb{A}_1+\mathbb{A}_2&=&4 c_1 \left(\sqrt{r(r+1)}-\sqrt{r_2(r_2+1)}\right),\\ 
\mathbb{A}_1-\mathbb{A}_2&=&-4c_1(r_2-r)
\end{eqnarray}
which vanish identically at $r=r_2$. 
\\
\indent The explicit form of the Killing spinor at the endpoint $r=r_2$ is given by 
\begin{equation}\label{SO(32)_SO(2)xSO(2)_eta_Soln}
\eta= Y_0e^{\frac{i z }{2 l}}r_2^{\frac{1}{5}} \begin{footnotesize}\begin{pmatrix} \sqrt{1+s} \\ -\sqrt{1-s} \end{pmatrix}\end{footnotesize}.
\end{equation}
We note that, in this case, there is no limit in which the $z$-circle decompactifies, so we expect the solution with be holographically dual to a four-dimensional $N=1$ SCFT. 

\subsection{Uplift to eleven dimensions}
As previously mentioned, the maximal gauged supergravity with $CSO(p,q,5-p-q)$ gauge group can be embedded in eleven-dimensional supergravity via a consistent truncation on $H^{p,q}\times T^{5-p-q}$. However, the complete truncation ansatz is presently known only for the case of $H^{5,0}$ or $S^4$. We will now uplift the $AdS_5\times \Sigma$ solution found in $SO(5)$ gauge group to eleven dimensions.
\\
\indent From the consistent $S^4$ truncation of eleven-dimensional supergravity given in \cite{M-theory_on_S4_1,M-theory_on_S4_2}, the metric can be written in notation of \cite{Pope_S3_S4} as
\begin{equation}
ds^2_{11}=\Delta^{\frac{1}{3}}ds^2_7+\frac{1}{g^2}\Delta^{-\frac{2}{3}}T^{-1}_{MN}D\mu^MD\mu^N
\end{equation}
in which $\mu^M$, $M=1,2,\ldots,5$ are coordinates on $S^4$ with $\mu^M\mu^M=1$. $T_{MN}$ is the $SL(5)$ scalar matrix obtained from the $SL(5)/SO(5)$ coset representative ${\mc{V}_M}^A$ as $T_{MN}={\mc{V}_M}^A{\mc{V}_N}^A$. The warp factor and the covariant derivative of $\mu^M$ are defined by
\begin{equation}
\Delta=T_{MN}\mu^M\mu^N\qquad \textrm{and}\qquad D\mu^M=d\mu^M+gA^{MN}\mu^N\, .
\end{equation}
In all of the solutions given here, the seven-dimensional three-form fields vanish, so the eleven-dimensional four-form field strength is simply given by
\begin{eqnarray}
F_{(4)}&=&-\frac{\Delta^{-2}U}{4!g^3}\epsilon_{M_1M_2M_3M_4M_5}\mu^{M_1}D\mu^{M_2}\wedge D\mu^{M_3}\wedge D\mu^{M_4}\wedge D\mu^{M_5}\nonumber \\
& &+\frac{1}{3!g^3}\Delta^{-2}\epsilon_{M_1M_2M_3M_4M_5}\mu^M\mu^NT^{M_1M}DT^{M_2N}\wedge D\mu^{M_3}\wedge D\mu^{M_4}\wedge D\mu^{M_5}\nonumber \\
& &+\frac{1}{4g^2}\Delta^{-1}\epsilon_{M_1M_2M_3M_4M_5}F^{M_1M_2}\wedge D\mu^{M_3}\wedge D\mu^{M_4} T^{M_5N}\mu^N
\end{eqnarray} 
with
\begin{eqnarray}
U&=&2T_{MN}T_{NP}\mu^M\mu^P-\Delta T_{MM}\nonumber \\
\textrm{and}\qquad DT^{MN}&=&dT^{MN}+gA^{MP}T^{PN}+gA^{NP}T^{MP}\, .
\end{eqnarray}
\indent By choosing the $S^4$ coordinates as
\begin{eqnarray}
& &\mu^1=\cos\beta\sin\alpha\cos\varphi,\qquad \mu^2=\cos\beta\sin\alpha\sin\varphi,\qquad \mu^3=\cos\beta\cos\alpha\cos\vartheta,\nonumber \\
& &\mu^4=\cos\beta\cos\alpha\sin\vartheta,\qquad \mu^5=\sin\beta\label{S4_coor}
\end{eqnarray}
with $\varphi\in [0,2\pi]$, $\vartheta\in [0,2\pi]$, $\alpha\in \left(0,\frac{\pi}{2}\right)$ and $\beta\in \left[-\frac{\pi}{2},\frac{\pi}{2}\right]$, the eleven-dimensional metric is given by
\begin{eqnarray}
ds^2_{11} &=& \frac{64 \Delta^{\frac{1}{3}}B r^{\frac{1}{20}}}{g^2 \left(r-\lambda\right)^{\frac{1}{4}}}\left[ds^2_{AdS_5}+\frac{r^{-\frac{5}{4}}}{64W\left(r-\lambda\right)^{\frac{7}{4}}}dr^2+\frac{g^2c_1{}^2W}{B}dz^2\right] \nonumber \\
&&\quad + \frac{1}{g^2}\Delta^{-\frac{2}{3}} \left\{ \left[e^{4(\phi_1+\phi_2)}\cos^2\beta + (e^{-2\phi_2}\cos^2\alpha + e^{-2\phi_1}\sin^2\alpha)\sin^2\beta\right]d\beta^2\right. \nonumber \\
&&+ \cos^2\beta(e^{-2\phi_1}\cos^2\alpha + e^{-2\phi_2}\sin^2\alpha)d\alpha^2 \nonumber \\
&&+ e^{-\phi_1-\phi_2}\sin 2\alpha\sin 2\beta\sinh(\phi_1-\phi_2)d\alpha d\beta \nonumber \\
&&+ e^{-2\phi_1}\cos^2\beta\sin^2\alpha(d\varphi - g\mathbb{A}_1 dz)^2 \nonumber \\
&&\left. + e^{-2\phi_2}\cos^2\beta\cos^2\alpha(d\vartheta - g\mathbb{A}_2 dz)^2 \right\}
\end{eqnarray}
with 
\begin{equation}
\Delta=e^{2\phi_1}\cos^2\beta\sin^2\alpha+e^{2\phi_2}\cos^2\beta\cos^2\alpha+e^{-4(\phi_1+\phi_2)}\sin^2\beta\, .
\end{equation}
For convenience, we also recall that 
\begin{eqnarray}
\phi_1&=&\frac{1}{2}\ln\left[r^{-\frac{1}{10}}(\sqrt{r}+\sqrt{r-1})\right],\nonumber \\
\phi_2&=&-\frac{1}{2}\ln\left[r^{\frac{1}{10}}(\sqrt{r}+\sqrt{r-1})\right]
\end{eqnarray}
with $\mathbb{A}_1$ and $\mathbb{A}_2$ given in \eqref{A121} and \eqref{A122}.
\\
\indent For $\phi_1=\phi_2=0$, the internal space becomes $S^4$ with the metric
\begin{equation}
ds^2_{S^4}=d\beta^2+\cos^2\beta(d\alpha^2+\sin^2\alpha d\varphi^2+\cos^2\alpha d\vartheta^2)
\end{equation}
and the volume given by
\begin{equation}
\textrm{vol}(S^4)=\int_{-\frac{\pi}{2}}^{\frac{\pi}{2}}\int_0^{\frac{\pi}{2}}\cos\alpha\sin\alpha\cos^3\beta d\alpha d\beta\int_{0}^{2\pi}\int_0^{2\pi}d\varphi d\vartheta=\frac{8\pi^2}{3}\, .
\end{equation}
\indent The four-form field strength is given by
\begin{eqnarray}
F_{(4)}&=&-\frac{U}{4!g^3\Delta^2}\epsilon_{M_1M_2M_3M_4M_5}\mu^{M_1}D\mu^{M_2}\wedge D\mu^{M_3}\wedge D\mu^{M_4}\wedge D\mu^{M_5}\nonumber \\
& &+\frac{\Delta^{-1}}{g^2}\left\{\cos^2\alpha\cos\beta\left(e^{2\phi_2}\cos^2\beta+e^{-4(\phi_1+\phi_2)}\sin^2\beta\right)F^{12}\wedge d\vartheta\wedge d\beta\right.\nonumber \\
& &+\sin^2\alpha\cos\beta\left(e^{2\phi_1}\cos^2\beta+e^{-4(\phi_1+\phi_2)}\sin^2\beta\right)F^{34}\wedge d\varphi\wedge d\beta\nonumber \\
& &\left.+e^{-4(\phi_1+\phi_2)}\cos\alpha\sin\alpha\cos^2\beta\sin\beta d\alpha\wedge(F^{12}\wedge d\vartheta+F^{34}\wedge d\varphi)\right\}\nonumber \\
& &
\end{eqnarray} 
with $F^{12}=\mathbb{A}'_1dr\wedge dz$, $F^{34}=\mathbb{A}'_2dr\wedge dz$ and
\begin{eqnarray}
U&=&e^{-8(\phi_1+\phi_2)}\sin^2\beta-e^{-4(\phi_1+\phi_2)}\left[2\sin^2\beta(e^{2\phi_1}+e^{2\phi_2})\right.\nonumber \\
& &\left.+\cos^2\beta(2e^{6(\phi_1+\phi_2)}+e^{2\phi_2}\cos^2\alpha+e^{2\phi_1}\sin^2\alpha)\right].
\end{eqnarray}
\indent We now proceed to the flux quantization condition
\begin{equation}
\frac{1}{(2\pi \ell_p)^3}\int F_{(4)}=N\in \mathbb{Z}\, .
\end{equation}
We first consider the integration along $S^4$ of the form
\begin{equation}
\int_0^{2\pi}\int_0^{2\pi}\int_{-\frac{\pi}{2}}^{\frac{\pi}{2}}\int_0^{\frac{\pi}{2}}\frac{U}{\Delta^2}\cos\alpha\sin\alpha\sin^2\beta d\alpha d\beta d\varphi d\vartheta=-2(2\pi)^2
\end{equation}
which gives
\begin{equation}
N=\frac{1}{\pi g^3\ell_p^3}\, .\label{quantize1}
\end{equation}
We then consider the integration along the four-cycle parametrized by $(rz\vartheta\beta)$ at $\alpha=0$ of the form
\begin{eqnarray}
\int_{-\frac{\pi}{2}}^{\frac{\pi}{2}}\int_0^{2\pi}\int_0^{2\pi}\int_1^{r_1}\frac{\cos\beta}{g^2\Delta}\left(e^{2\phi_2}\cos^2\beta+e^{-4(\phi_1+\phi_2)}\sin^2\beta\right)\mathbb{A}'_1dr dz d\vartheta d\beta & &\nonumber \\
=(2\pi\ell_p)^3M\quad & &
\end{eqnarray}
for $M\in \mathbb{Z}$. After using explicit solutions for $\phi_1$, $\phi_2$ and $\mathbb{A}_1$ as well as \eqref{quantize1}, we eventually find 
\begin{equation}
\frac{1-r_1+\sqrt{r_1(r_1-1)}}{4r_1-3}=\frac{Ml}{N}\, .
\end{equation}
This can be used to express $r_1$ in terms of integral fluxes $M$ and $N$ as
\begin{equation}
r_1=\frac{N^2+7NMl+6M^2l^2\pm N\sqrt{N^2-2NMl-3M^2l^2}}{8Ml(N+Ml)}\, .
\end{equation}
\indent At this point, we note again that if we compute the holographic central charge of the dual $N=1$ SCFT, the result becomes infinite since the integral over the internal six-dimensional manifold diverges at $r=1$. As previously mentioned, at $r=1$, the $S^1$ parametrized by $z$ decompactifies, and the seven-dimensional metric is asymptotic to the $AdS_7$ geometry. Accordingly, the solution should be interpreted as describing a codimension-2 conformal defect within the six-dimensional $N=(2,0)$ SCFT as in \cite{Gutperle_coD2_defect}.
\section{$SO(2)_{\text{diag}}$ symmetric solutions}\label{Y_SO(2)D_Sec}
We now look for supersymmetric $AdS_5\times \Sigma$ solutions with $SO(2)_{\text{diag}}$ residual symmetry of the $CSO(p,q,5-p-q)$ gauge groups. In this case, there are four singlet scalars, and the $SL(5)/SO(5)$ coset representative takes the form
\begin{equation}
\mc{V}=e^{\phi_1\hat{Y}_1+\phi_2\hat{Y}_2+\phi_3\hat{Y}_3+\phi_4\hat{Y}_4}
\end{equation}
with $\hat{Y}_1$ and $\hat{Y}_2$ given in \eqref{Y12_def}. The other two additional non-compact generators are given by
\begin{equation}
\hat{Y}_3=e_{1,3}+e_{3,1}+e_{2,4}+e_{4,2}\qquad \textrm{and}\qquad
\hat{Y}_4=e_{1,4}+e_{4,1}-e_{2,3}-e_{3,2}\, .
\end{equation}
\indent Unlike the previous case, in order to implement the $SO(2)_{\textrm{diag}}$ symmetry generated by $X_{12}+\kappa X_{34}$ the two $SO(2)\times SO(2)$ gauge fields are now related by 
\begin{equation}
A^{12}=\kappa A^{34}\, .\label{U1_diag_con}
\end{equation}
However, solving the gauge field equations leads to the conditions that the scalars $\phi_3$ and $\phi_4$ must vanish in order to find non-trivial solutions.
\\ 
\indent With $\phi_3=\phi_4=0$, the analysis of the BPS equations follows essentially the same line as in the previous section. We will not repeat all the details here but only give the main results leading to supersymmetric $AdS_5\times \Sigma$ solutions. To implement the condition \eqref{U1_diag_con}, we will set $a_2=\kappa a_1$, and the condition given in \eqref{1st_Strongest_con} now takes the form
\begin{equation}
a_1\lambda  e^{2 (\phi_1+\phi_2)} \left(e^{2 \phi_2}-\kappa  e^{2 \phi_1}\right)=0\, .\label{SO(2)d_Strongest_con}
\end{equation}
For gauge groups with $\lambda=0$, this condition is identically satisfied, but we do not find any solutions to the remaining BPS equations in this case. On the other hands, for $\lambda\neq0$, equation \eqref{SO(2)d_Strongest_con} admits real solutions for $\phi_1$ and $\phi_2$ only for $\kappa=1$ giving rise to $\phi_2=\phi_1$. Accordingly, we will consider only $SO(5)$ ($\kappa=\lambda=1$) and $SO(4,1)$ ($\kappa=-\lambda=1$) gauge groups in this case. 
\\
\indent Setting $\kappa=1$, $a_2=a_1$, and $\phi_2=\phi_1$, the solutions for $f$ and $h_1$ obtained from \eqref{f_gen_sol} and \eqref{h1_gen_sol} become
\begin{eqnarray}
f&=&\frac{\sqrt{a_1} e^{4 \phi_1}}{\sqrt{s \left(e^{10 \phi_1}-\lambda \right)}},\label{f_SO(2)d_sol}\\
h_1&=&\frac{400 \sqrt{a_1} e^{24 \phi_1} (\phi'_1)^2}{\left(e^{10 \phi_1}-\lambda \right)^2 \left[\sqrt{a_1} g^2 e^{20 \phi_1}-64 \sqrt{s \left(e^{10 \phi_1}-\lambda \right)}\right]}\label{h1_SO(2)d_sol}
\end{eqnarray}
while the differential equation \eqref{gen_A_prime} reduces to
\begin{equation}
\widehat{\mathbb{A}}'=\frac{40 \widehat{\mathbb{A}} \phi'_1}{3 \lambda  e^{10 \phi_1}-4}\, .\label{SO(2)d_A_prime}
\end{equation}
The solution to this equation takes the form of
\begin{equation}
\widehat{\mathbb{A}}=c_1 \left(3 \lambda -4 e^{-10 \phi_1}\right)
\end{equation}
where $c_1$ is an integration constant. Using the definition \eqref{AandA}, we then find 
\begin{equation}
\mathbb{A}_2=\mathbb{A}_1=\frac{c_1}{2} \left(3 \lambda -4 e^{-10 \phi_1}\right)-\frac{q}{g}\, .
\end{equation}
\indent With all these, the solution for $h_2$ can be obtained from \eqref{sqrth1h1_1} as
\begin{equation}
h_2=\frac{c_1{}^2 e^{-16 \phi_1} \left[\sqrt{a_1} g^2 e^{20 \phi_1}-64 \sqrt{s \left(e^{10 \phi_1}-\lambda \right)}\right]}{\sqrt{s \left(e^{10 \phi_1}-\lambda \right)}}\, .
\end{equation}
It can also be verified that all the remaining BPS equations and the field equations are satisfied if 
\begin{equation}\label{SO(2)d_sign_Con}
s \left(e^{10 \phi_1}-\lambda \right)>0\, .
\end{equation}
\indent Finally, the solution for the two-component spinor $\eta$ can be determined to be  
\begin{equation}
\eta=e^{iqz}Ye^{5\phi_1}\begin{pmatrix} \sqrt{g\lambda(a_1)^{\frac{1}{4}}+8se^{-10\phi_1}\left[s(e^{10\phi_1}-\lambda)\right]^{\frac{1}{4}}} \\ -\sqrt{g\lambda(a_1)^{\frac{1}{4}}-8se^{-10\phi_1}\left[s(e^{10\phi_1}-\lambda)\right]^{\frac{1}{4}}}\end{pmatrix}.
\end{equation}
The function $Y$ can be obtained from solving \eqref{partial_r_con}. The result is given by
\begin{equation}
Y=\frac{Y_0\,e^{-4\phi_1}}{\left(e^{10\phi_1}-\lambda\right)^{\frac{1}{8}}}
\end{equation}
with $Y_0$ being an integration constant. 
\\
\indent To further analyse the solution, we again define the following parameter
\begin{equation}\label{para_B_def}
B=\frac{g^2\sqrt{a_1}}{64}
\end{equation}
together with the function
\begin{equation}\label{Def_W}
W=B-e^{-20\phi_1}\sqrt{s(e^{10\phi_1}-\lambda)}\, .
\end{equation}
In terms of these quantities, the seven-dimensional metric reads
\begin{equation}
ds^2_7=\frac{64B\,e^{4\phi_1}}{g^2\sqrt{s(e^{10\phi_1}-\lambda)}}ds^2_{AdS_5}+\frac{400B\,e^{4\phi_1}\,(\phi'_1)^2}{g^2W(e^{10\phi_1}-\lambda)^2}dr^2+\frac{64c_1{}^2We^{4\phi_1}}{\sqrt{s(e^{10\phi_1}-\lambda)}}dz^2\, .
\end{equation}
As in the previous section, it turns out that the resulting solution takes the same form as the $AdS_5\times \Sigma$ solution with $SO(2)_R$ symmetry from seven-dimensional $N=2$ gauged supergravity coupled to three vector multiplets studied in \cite{7D_N2_disk}. This can be seen by the following identification 
\begin{equation}
\phi_1=\frac{\sigma}{4},\qquad g=32h,\qquad \lambda=m=\frac{\widetilde{g}_1}{16h},\qquad c_1=\frac{\mathcal{C}}{16h}\, .
\end{equation}
The $SO(2)_R$ gauge field $A_1$ in \cite{7D_N2_disk} is identified as the $SO(2)_{\text{diag}}$ gauge field in the present case via 
\begin{equation}
A_1=-2\mathbb{A}_d=-2\left(\mathbb{A}_1+\mathbb{A}_2\right).
\end{equation}
In more detail, the solutions in $SO(5)$ ($\lambda=1$) and $SO(4,1)$ ($\lambda=-1$) gauge groups found in this work are respectively equivalent to the $SO(2)_R$ symmetric solutions in $SO(4)$ gauge group with $\widetilde{g}_1=16h$ and $SO(3,1)$ gauge group with $\widetilde{g}_1=-16h$ studied in \cite{7D_N2_disk}. As pointed out in the previous section, all of the present solutions obtained from the maximal gauged supergravity have clear higher-dimensional origin while only the solution from $SO(4)$ gauge group with $\widetilde{g}_1=16h$ studied in \cite{7D_N2_disk} can be embedded in eleven dimensions.  
\\
\indent To discuss the regularity conditions, we fix the solution of $\phi_1$ to be
\begin{equation}
\phi_1=-\frac{1}{10}\ln r\, .
\end{equation}
In terms of the radial coordinate $r$, the seven-dimensional metric reads
\begin{eqnarray}
ds^2_7=\frac{64B\,r^{\frac{1}{10}}}{g^2\sqrt{s(1-\lambda r)}}\left[ds^2_{AdS_5}+ds^2_\Sigma\right]\nonumber\\
\end{eqnarray}
with 
\begin{eqnarray}
ds^2_\Sigma=\frac{r^{-\frac{1}{2}}}{16W\left[s(1-\lambda r)\right]^{\frac{3}{2}}}dr^2+\frac{g^2c_1{}^2W}{B}dz^2\label{Sigma_Met}
\end{eqnarray}
and
\begin{equation}
W=B-r^{\frac{3}{2}}\sqrt{s(1-\lambda r)}\, .
\end{equation}
The equation $W=0$ admits four roots given by
\begin{equation}\label{r_Roots}
r_{1\pm}=\frac{1}{4}\left[\lambda+2\sqrt{X}\pm\sqrt{\left(3-4X\right)+\frac{\lambda}{\sqrt{X}}}\,\right]
\end{equation}
and
\begin{equation}\label{r2_Roots}
r_{2\pm}=\frac{1}{4}\left[\lambda-2\sqrt{X}\pm\sqrt{\left(3-4X\right)-\frac{\lambda}{\sqrt{X}}}\,\right]
\end{equation}
with
\begin{eqnarray}
X&=&\frac{1}{4}+\left(\frac{B}{3\sqrt{2}}\right)^{\frac{2}{3}}\lambda  s\left(9-\sqrt{81-768 B^2 \lambda s}\right)^{\frac{1}{3}}\nonumber \\
& &+\left(\frac{B}{3\sqrt{2}}\right)^{\frac{2}{3}}\lambda  s\left(9+\sqrt{81-768 B^2 \lambda s}\right)^{\frac{1}{3}}\, .\label{X_def}
\end{eqnarray}
\indent The $SO(2)_{\text{diag}}$ gauge field and the two-component spinor $\eta$ take the form 
\begin{eqnarray}
\mathbb{A}_d&=&c_1\left(3\lambda-4r\right)-\frac{2q}{g}\label{SO(2)d_soln}\\
\text{and}\qquad\eta&=&e^{iqz}Y_0\frac{2^{3/2}r^{\frac{1}{40}}}{\left[s\left(1-\lambda r\right)\right]^{\frac{1}{8}}}\begin{pmatrix} \sqrt{\sqrt{B}+sr^{\frac{3}{4}}\left[s(1-\lambda r)\right]^{\frac{1}{4}}} \\ -\sqrt{\sqrt{B}-sr^{\frac{3}{4}}\left[s(1-\lambda r)\right]^{\frac{1}{4}}} \end{pmatrix}.\qquad\label{eta_Soln}
\end{eqnarray}
We now discuss possible ranges of the radial coordinate $r$ for obtaining regular solutions together with the corresponding behaviors of the solutions. Some parts of the analysis have also been performed in \cite{7D_N2_disk}. In this paper, we will discuss this in detail including the uplifted solutions to eleven-dimensional supergravity in the case of $SO(5)$ gauge group. Unlike the previous solution with $SO(2)\times SO(2)$ symmetry, there exist a number of coordinate ranges for which regular solutions are possible.

\subsection{Regular solutions in $SO(5)$ gauge group}\label{Y_SO(2)D_SO(5)_Sec} 
For $SO(5)$ gauge group with $\lambda=1$, we find a number of possibilities for both $s=+1$ and $-1$.
\subsubsection{Solutions with $s=+1$}
\begin{figure}[h!]
  \centering
  \begin{subfigure}[b]{0.326\linewidth}
    \includegraphics[width=\linewidth]{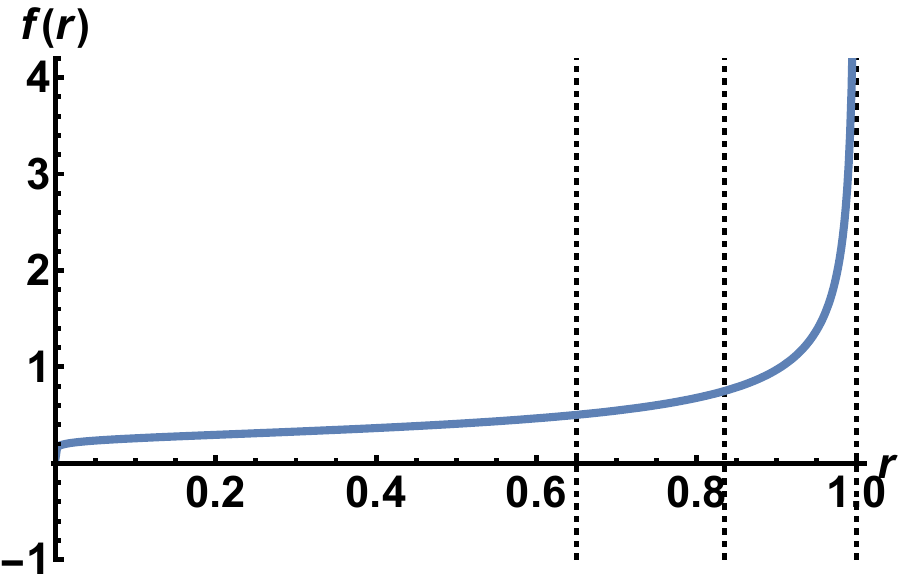}
\caption{$f$ solution}
  \end{subfigure}
  \begin{subfigure}[b]{0.326\linewidth}
    \includegraphics[width=\linewidth]{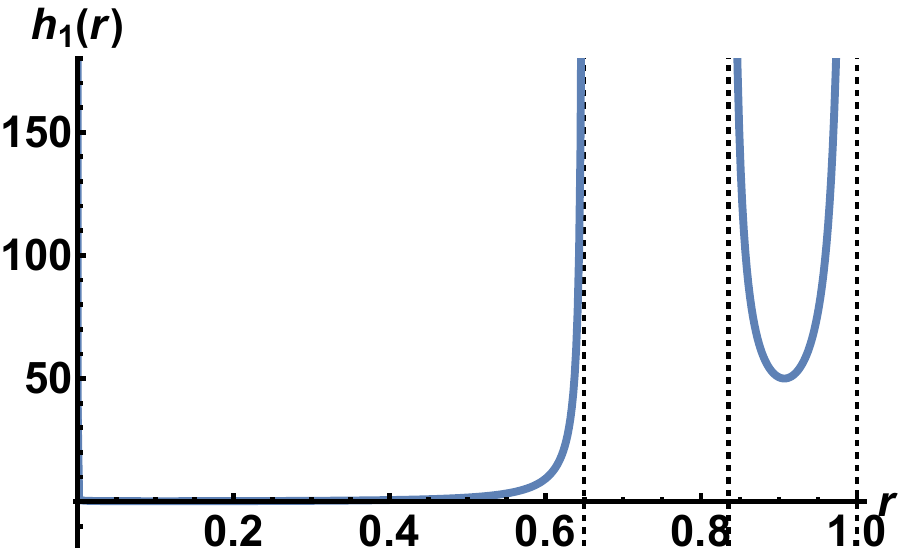}
\caption{$h_1$ solution}
  \end{subfigure}
  \begin{subfigure}[b]{0.326\linewidth}
    \includegraphics[width=\linewidth]{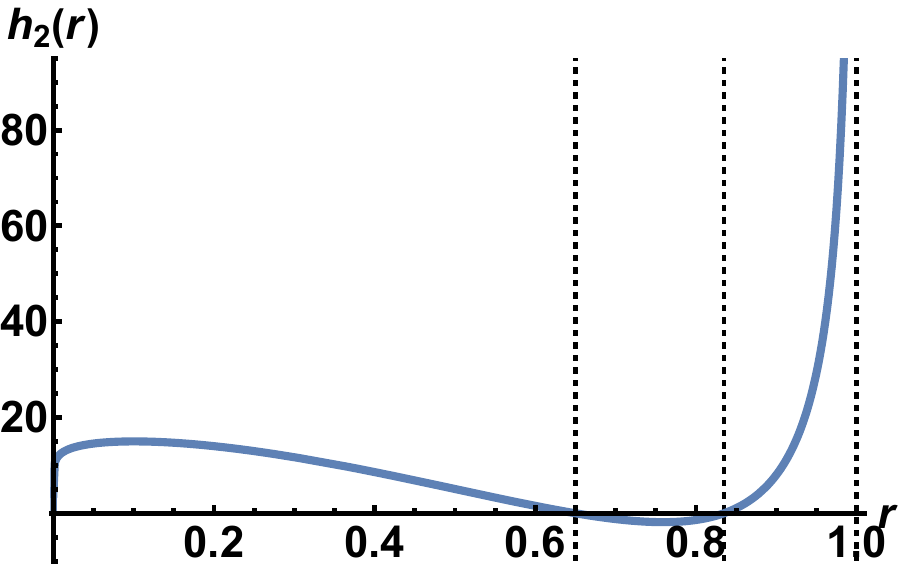}
\caption{$h_2$ solution}
  \end{subfigure}
  \caption{Numerical plots of the warp factors for the $SO(2)_{\text{diag}}$ symmetric solution from $SO(5)$ gauge group, evaluated at $s=+1$, $B=0.31$, $c_1=1$, and $g=8$. The solution is regular within the intervals $0<r<r_{1-}=0.65$ and $r_{1+}=0.84<r<1$. The three vertical dashed lines mark the boundaries of these regions excluding $r = 0$.}
  \label{Fig1}
\end{figure}

\begin{figure}[h!]
  \centering
  \begin{subfigure}[b]{0.326\linewidth}
    \includegraphics[width=\linewidth]{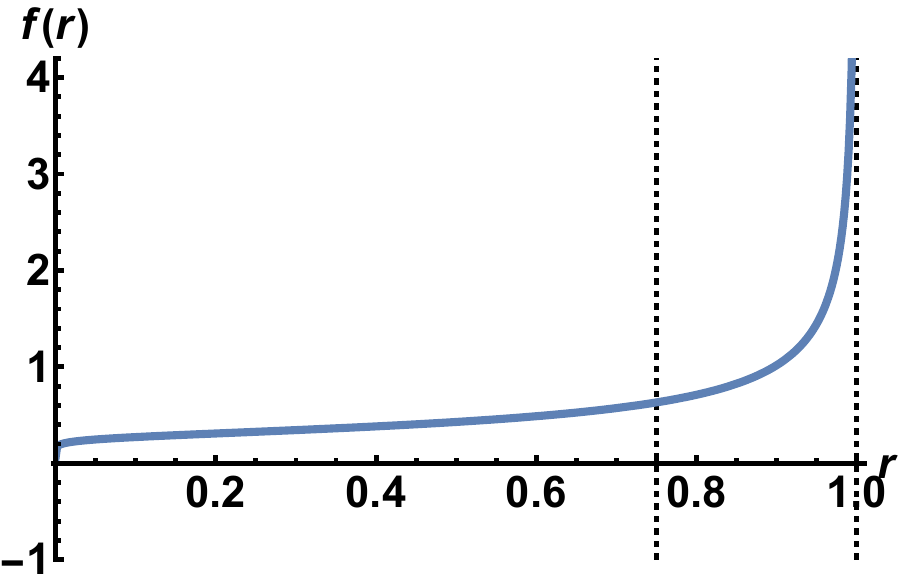}
\caption{$f$ solution}
  \end{subfigure}
  \begin{subfigure}[b]{0.326\linewidth}
    \includegraphics[width=\linewidth]{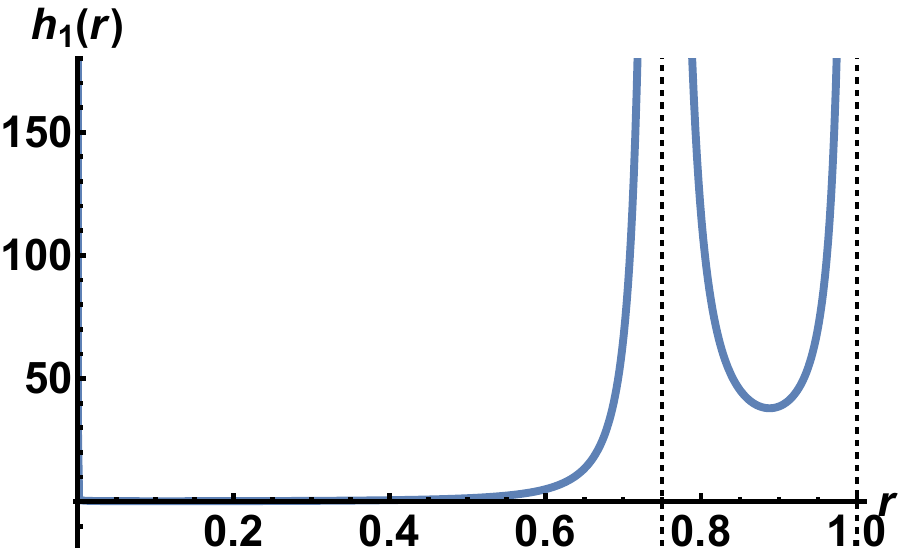}
\caption{$h_1$ solution}
  \end{subfigure}
  \begin{subfigure}[b]{0.326\linewidth}
    \includegraphics[width=\linewidth]{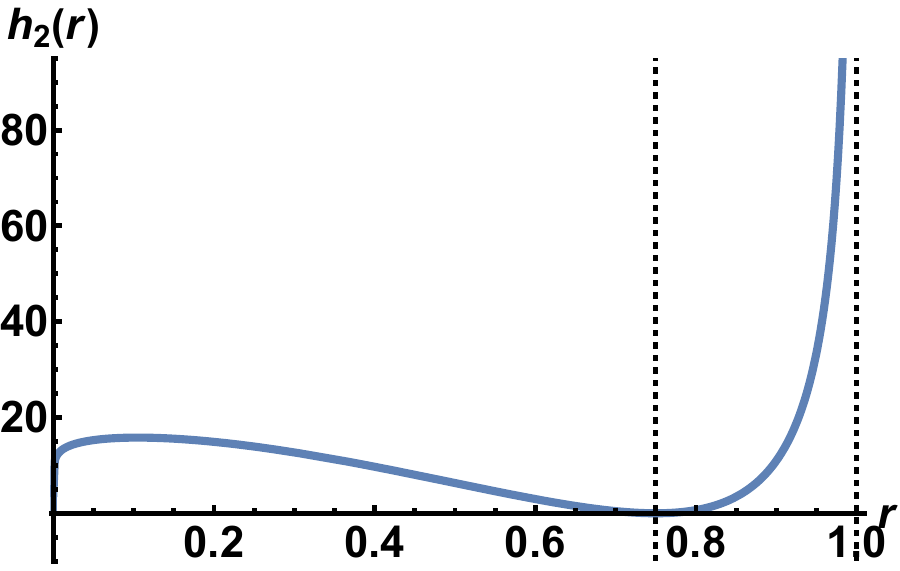}
\caption{$h_2$ solution}
  \end{subfigure}
  \caption{Numerical plots of the warp factors for the $SO(2)_{\text{diag}}$ symmetric solution from $SO(5)$ gauge group, evaluated at $s=+1$, $B=\frac{3\sqrt{3}}{16}$, $c_1=1$, and $g=8$. The solution is regular within the intervals $0<r<r_{1-}=0.75$ and $r_{1+}=0.75<r<1$. The two vertical dashed lines mark the boundaries of these regions excluding $r = 0$.}
  \label{Fig2}
\end{figure}

\begin{figure}[h!]
  \centering
  \begin{subfigure}[b]{0.326\linewidth}
    \includegraphics[width=\linewidth]{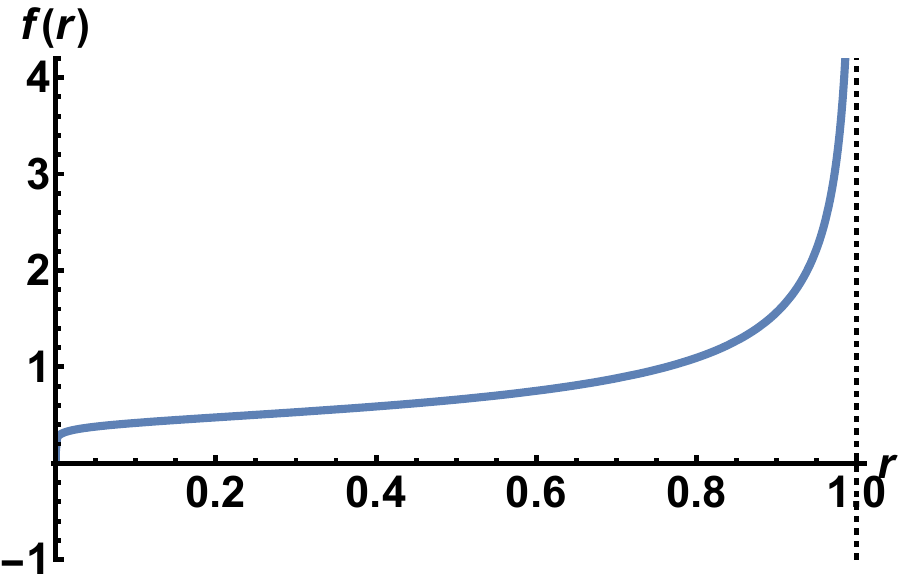}
\caption{$f$ solution}
  \end{subfigure}
  \begin{subfigure}[b]{0.326\linewidth}
    \includegraphics[width=\linewidth]{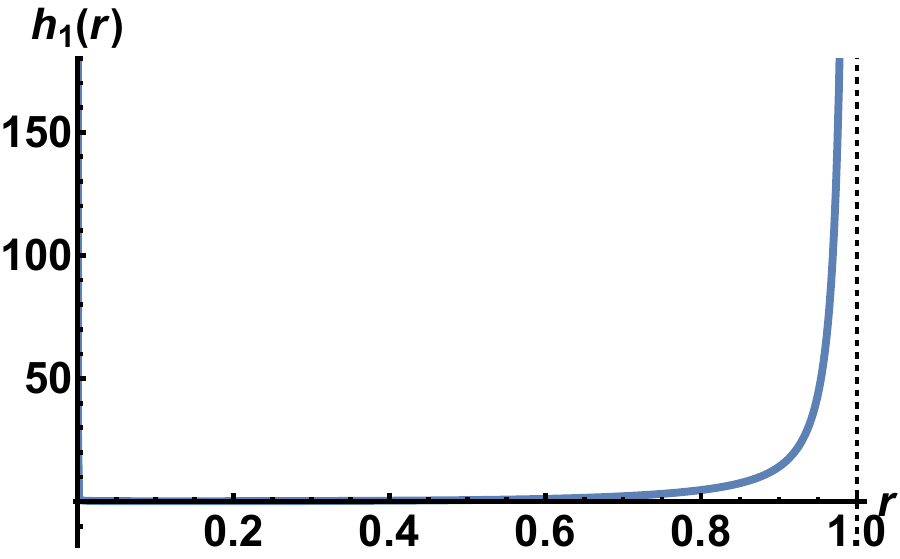}
\caption{$h_1$ solution}
  \end{subfigure}
  \begin{subfigure}[b]{0.326\linewidth}
    \includegraphics[width=\linewidth]{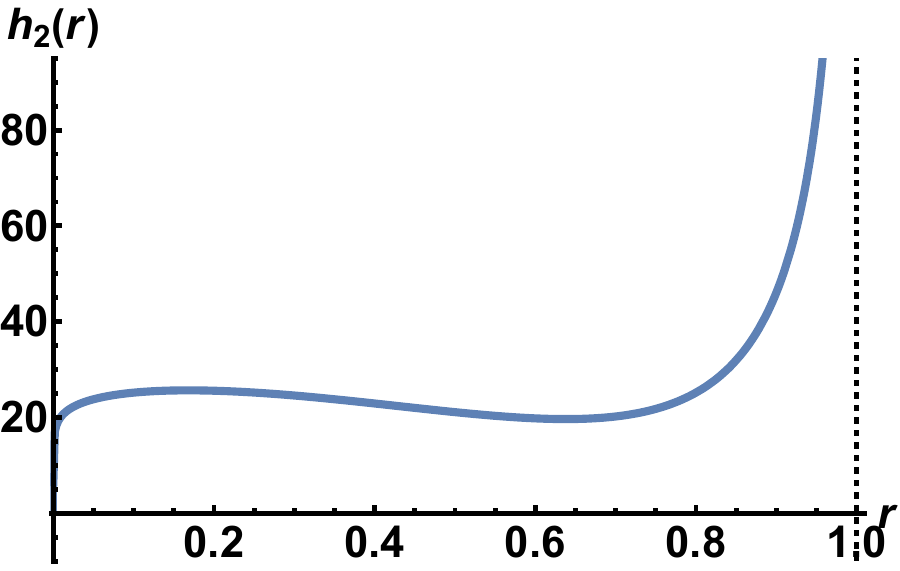}
\caption{$h_2$ solution}
  \end{subfigure}
  \caption{Numerical plots of the warp factors for the $SO(2)_{\text{diag}}$ symmetric solution from $SO(5)$ gauge group, evaluated at $s=+1$, $B=0.5$, $c_1=1$, and $g=8$. The solution is regular within the interval $0<r<1$. The vertical dashed line marks the boundary at $r=1$.}
  \label{Fig3}
\end{figure}

Possible ranges of $r$ for regular solutions with $s=+1$ exist between $r=0$ and $r=1$. Since $\lambda=s=+1$, we find that there are three different possibilities depending on the value of $\sqrt{81-768 B^2}$ in $r_{1\pm}$ and $r_{2\pm}$. For $B<\frac{3\sqrt{3}}{16}$ or $81-768 B^2<0$, we find two possible ranges
\begin{equation}
0<r<r_{1-}\qquad\text{ and }\qquad r_{1+}<r<1. \label{2_pos_r}
\end{equation}
For the special case with $B=\frac{3\sqrt{3}}{16}$, there are still two possible ranges of $r$ as given in \eqref{2_pos_r} but with $r_{1-}=r_{1+}$. For $B>\frac{3\sqrt{3}}{16}$, there is only one possible range of $r$
\begin{equation}
0<r<1\, .
\end{equation}
Examples of numerical plots for the three warp factors for $c=1$ and $g=8$ in these three cases with $B=0.31, \frac{3\sqrt{3}}{16},$ and $0.5$ are respectively given in figures \ref{Fig1}, \ref{Fig2}, and \ref{Fig3}. 

\begin{figure}[h!]
  \centering
    \includegraphics[width=0.6\linewidth]{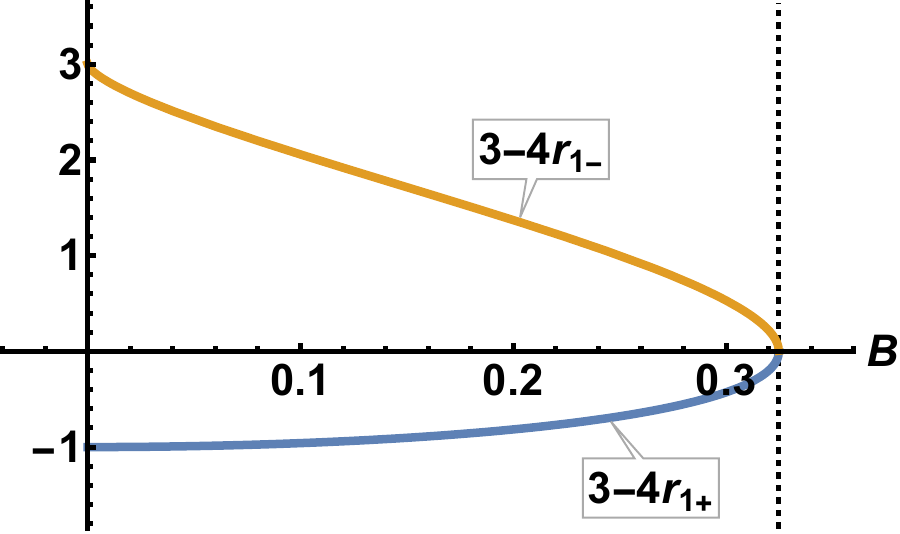}
  \caption{Numerical plot of the functions $3-4r_{1-}$ (orange line) and $3-4r_{1+}$ (blue line) appearing in condition \eqref{s+_smooth_con} with $s=+1$ in $SO(5)$ gauge group. Both functions approach zero as $B\rightarrow\frac{3\sqrt{3}}{16}$ (the vertical dashed line).}
  \label{Fig4}
\end{figure}

As $r\rightarrow0$, the seven-dimensional metric again becomes conformal to a product of $AdS_5$ and a cylinder. With the new radial coordinate $\rho$ given by $r=3^{\frac{4}{3}}B^{\frac{2}{3}}c_1{}^{\frac{4}{3}}g^{\frac{4}{3}} \rho^{\frac{4}{3}}$, the metric near $\rho=0$ is given by
\begin{equation}\label{SO(2)d_rto0_metric}
ds_7^2\approx\frac{64(3B^8c_1 g \rho){}^{\frac{2}{15}}}{g^2}\left[ds^2_{AdS_5}+c_1{}^2 g^2(d\rho^2+dz^2)\right].
\end{equation}
We also note that the warp factor of the $AdS_5$ factor vanishes in this limit leading to a singularity. This can be seen from the scalar curvature of the metric which takes the form
\begin{equation}
R\sim \frac{g^2}{r^{\frac{8}{5}}}
\end{equation}
near $r=0$. 
\\
\indent On the other hand, as $r\rightarrow 1$, the seven-dimensional metric also becomes conformal to a product of $AdS_5$ and a cylinder. With the new radial coordinate $R$ given by $r=1-B^2 c_1{}^4 g^4 R^4$, the metric near $R=0$ reads
\begin{equation}\label{SO(2)d_rto1_metric}
ds_7^2\approx\frac{64}{c_1{}^2 g^4 R^2}\left[ds^2_{AdS_5}+c_1{}^2 g^2(dR^2+dz^2)\right].
\end{equation}
Similar to the $SO(2)\times SO(2)$ symmetric solution, in this limit, the $z$-circle decompactifies, and the solution is asymptotic to $AdS_7$ geometry with $AdS_5\times S^1$ boundary. We again interpret this solution as describing a codimension-2 defect within the $N=(2,0)$ SCFT in six dimensions.
\\
\indent As $r\rightarrow r_{1\pm}$, we introduce a new radial coordinate $\rho=\frac{1}{2}\sqrt{r_{1\pm}-r}$ in terms of which the six-dimensional metric, as $r\rightarrow r_{1\pm}$ or $\rho\rightarrow0$, is approximately given by  
\begin{equation}\label{Case-1_spindle_metric}
ds_7^2\approx\frac{64 B r_{1\pm}^{\frac{1}{10}}}{g^2 \sqrt{1-r_{1\pm}}}\left[ds^2_{AdS_5}+\frac{d\rho^2+c_1{}^2 g^2 \rho ^2 \left(3-4 r_{1\pm}\right)^2dz^2}{-W'\left(r_{1\pm}\right)\sqrt{r_{1\pm}}\left(1-r_{1\pm}\right)^{\frac{3}{2}}}\right].
\end{equation}
In this case, the $z$-circle shrinks smoothly, giving rise to an $\mathbb{R}^2/\mathbb{Z}_l$ orbifold at $r=r_{1\pm}$ if we impose the condition
\begin{equation}\label{s+_smooth_con}
c_1= \frac{1}{l \left|g(3-4r_{1\pm})\right|},\qquad l=1,2,3,\ldots\, .
\end{equation}
We have used an absolute sign for the function $g(3-4r_{1\pm})$ since we want $c_1$ to be positive. Moreover, substituting $r_{1\pm}$ from \eqref{r_Roots} into this function, we find that $0<3-4r_{1-}<3$ and $-1<3-4r_{1+}<0$ in the range $0<B<\frac{3\sqrt{3}}{16}$ as seen from a numerical plot given in figure \ref{Fig4}. We also note that this should correspond to the surface plot given in figure 7 of \cite{7D_N2_disk} with $m=1$. From figure \ref{Fig4}, we can see that both functions approach zero as $B\rightarrow\frac{3\sqrt{3}}{16}$ (the vertical dashed line). Therefore, for the special case with $B=\frac{3\sqrt{3}}{16}$, the $z$-circle does not shrink smoothly at $r=r_{1-}=r_{1+}=0.75$ since $c_1$ diverges due to $g(3-4r_{1\pm})=0$.

As in the previous case, we can calculate the Euler characteristic of $\Sigma$ from the metric $ds^2_{\Sigma}$ given in \eqref{Sigma_Met}
\begin{eqnarray}
\chi(\Sigma)&=&\frac{1}{4\pi}\int_{\Sigma}R_{\Sigma}\text{vol}_{\Sigma}\ =\ \mp\frac{c_1 g \left(3-4 r_{1\pm}\right) \left(1-r_{1\pm}\right)^{\frac{1}{4}} r_{1\pm}^{\frac{3}{4}}}{\sqrt{B}}\nonumber \\
&=&\mp c_1 g\left(3-4 r_{1\pm}\right)\ =\ \frac{1}{l}\, .\label{SO(2)d_SO(5)gg_chi}
\end{eqnarray}
The integration has been performed on the interval $0<r<r_{1-}$ or $r_{1+}<r<1$, and $0<z<2\pi$. We have also used $B=r_{1\pm}^{\frac{3}{2}}\sqrt{(1-r_{1\pm})}$ obtained from $W(r_{1\pm})=0$. Therefore, the space $\Sigma$ with metric \eqref{Sigma_Met} forms two topological disks, one with an origin at $r = r_{1-}$ and a boundary at $r = 0$ and another one with an origin at $r = r_{1+}$ and a boundary at $r = 1$. At these boundaries, the $z$-circle does not collapse, and there is a $\mathbb{Z}_l$ orbifold singularity at the center of each disk. 
\\
\indent In order for the $SO(2)_{\text{diag}}$ gauge field to vanish at $r=r_{1\pm}$, we fix the constant $q$ to be
\begin{equation}
q=\frac{c_1 g}{2}\left(3-4 r_{1\pm}\right)
\end{equation}
leading to
\begin{equation}
\mathbb{A}_d=4 c_1 \left(r_{1\pm}-r\right).\label{SO(2)d_Ad_final_soln}
\end{equation}
The Killing spinor $\eta$ at $r= r_{1\pm}$ reads
\begin{equation}\label{SO(2)diag_SO(2)_s+_spindle_eta}
\eta=4 Y_0 e^{\mp\frac{i z }{2 l}} r_{1\pm}^{\frac{2}{5}}  \begin{pmatrix} 1 \\ 0 \end{pmatrix}.
\end{equation}
which is well-behaved near $r=r_{1\pm}$ and hence globally defined on the disk $\Sigma$. 

\subsubsection{Solutions with $s=-1$}
\begin{figure}[h!]
  \centering
  \begin{subfigure}[b]{0.326\linewidth}
    \includegraphics[width=\linewidth]{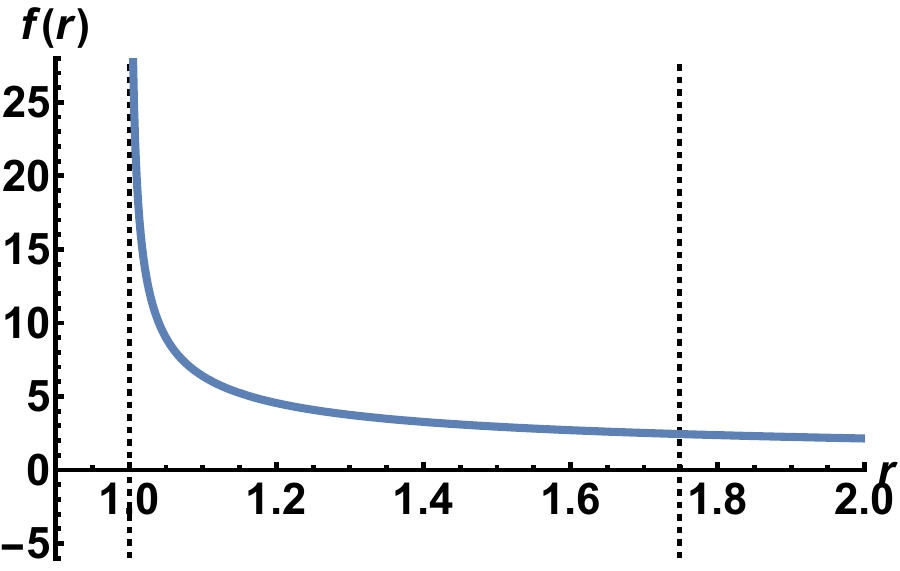}
\caption{$f$ solution}
  \end{subfigure}
  \begin{subfigure}[b]{0.326\linewidth}
    \includegraphics[width=\linewidth]{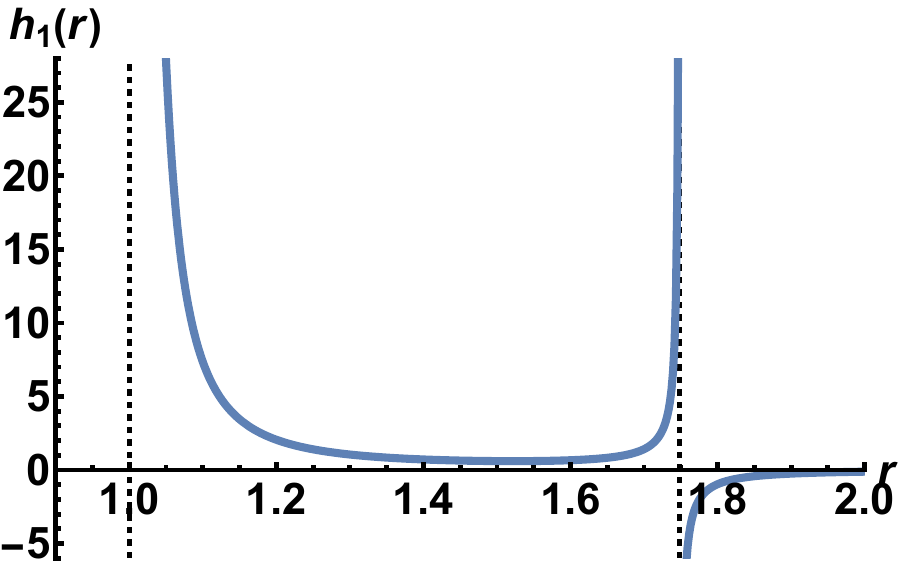}
\caption{$h_1$ solution}
  \end{subfigure}
  \begin{subfigure}[b]{0.326\linewidth}
    \includegraphics[width=\linewidth]{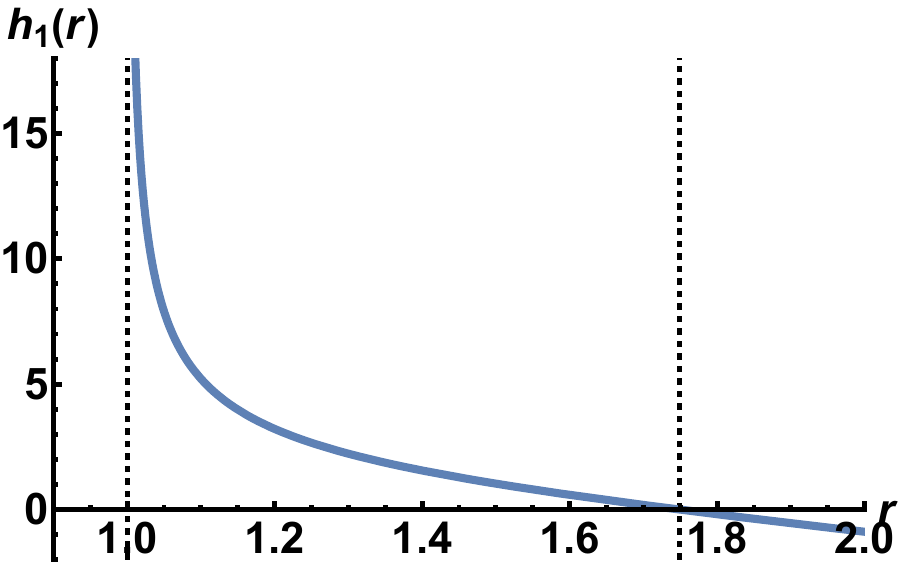}
\caption{$h_2$ solution}
  \end{subfigure}
  \caption{Numerical plots of the warp factors for the $SO(2)_{\text{diag}}$ symmetric solution from $SO(5)$ gauge group, evaluated at $s=-1$, $B=2$, $c_1=\frac{1}{8}$, and $g=8$. The solution is regular within the interval $1<r<r_{1+}=1.748$. The two vertical dashed lines mark the boundaries of these regions.}
  \label{Fig5}
\end{figure}

For $s=-1$, we find regular solutions only when $B>0$ and
\begin{equation}\label{SO(2)d_SO(5)gg_s-_Pos_Range}
1<r<r_{1+}\, .
\end{equation}
We plot a representative solution with $B=2$, $c=\frac{1}{8}$, and $g=8$ in figure \ref{Fig5}. As $r\rightarrow1$, setting $r=1+B^2 c_1{}^4 g^4 R^4$, we find the seven-dimensional metric near $R=0$ as given in \eqref{SO(2)d_rto1_metric}. Thus, the seven-dimensional space-time is again asymptotic to $AdS_7$ when $r\rightarrow1$ as in the previous cases.

\begin{figure}[h!]
  \centering
    \includegraphics[width=0.5\linewidth]{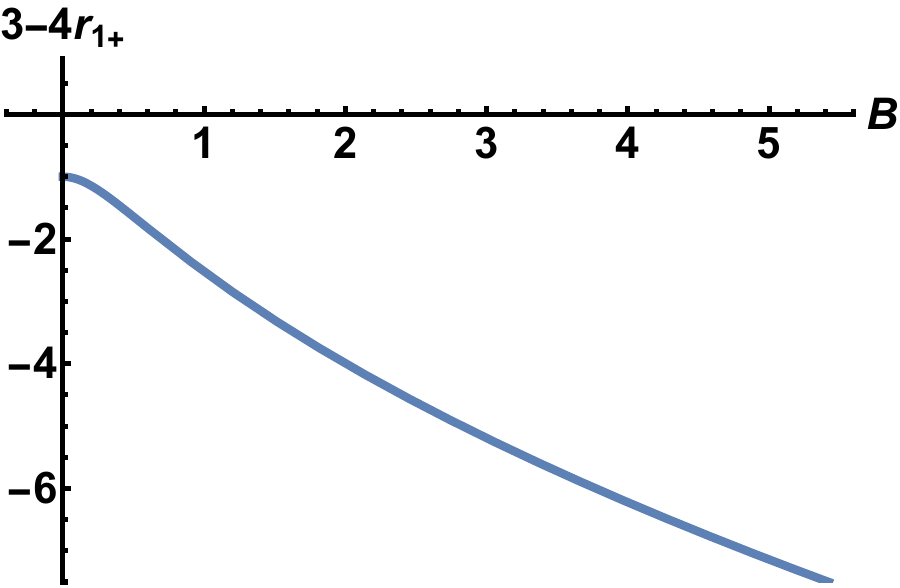}
  \caption{A numerical plot of the function $3-4r_{1+}$ appearing in the condition \eqref{Case_s-_smooth_con}  with $s=-1$ in $SO(5)$ gauge group. The function approaches $-1$ as $B\rightarrow 0$.}
  \label{Fig6}
\end{figure}

On the other end, as $r\rightarrow r_{1+}$, we find the following form of the seven-dimensional metric
\begin{equation}\label{Case_s-_spindle_metric}
ds_7^2\approx\frac{64 B r_{1+}^{\frac{1}{10}}}{g^2 \sqrt{r_{1+}-1}}\left[ds^2_{AdS_5}+\frac{d\rho^2+c_1{}^2 g^2 \rho ^2 \left(3-4 r_{1+}\right)^2dz^2}{-W'\left(r_{1+}\right)\sqrt{r_{1+}}\left(r_{1+}-1\right)^{\frac{3}{2}}}\right]
\end{equation}
after changing the radial coordinate to $\rho=\frac{1}{2}\sqrt{r_{1+}-r}$. The $z$-circle shrinks smoothly if we impose
\begin{equation}\label{Case_s-_smooth_con}
c_1= -\frac{1}{lg(3-4r_{1+})},\qquad l=1,2,3,\ldots\, .
\end{equation}
In this case, we find that the function $3-4r_{1+}$ is negative as can be seen from the numerical plot in figure \ref{Fig6}. With the condition \eqref{Case_s-_smooth_con}, the Euler characteristic of $\Sigma$ can be computed as in \eqref{SO(2)d_SO(5)gg_chi}. In this case, the integration is performed on the interval $1<r<r_{1+}$ and $0<z<2\pi$, and we also use $B=r_{1+}^{\frac{3}{2}}\sqrt{(r_{1+}-1)}$ obtained from $W(r_{1+})=0$.

Unlike all the previous cases with $s=+1$, only the lower component of the Killing spinor $\eta$ is now non-vanishing as $r\rightarrow r_{1+}$ 
\begin{equation}\label{Case_s-_eta}
\eta=-4 Y_0 e^{-\frac{i z }{2 l}} r_{1+}^{\frac{2}{5}}  \begin{pmatrix} 0 \\ 1 \end{pmatrix}\, .
\end{equation}
The $SO(2)_{\text{diag}}$ gauge field also takes the same form as in \eqref{SO(2)d_Ad_final_soln} and vanishes at $r=r_{1+}$. 

\subsection{Regular solutions in $SO(4,1)$ gauge group}\label{Y_SO(2)D_SO(4,1)_Sec} 
\begin{figure}[h!]
  \centering
  \begin{subfigure}[b]{0.326\linewidth}
    \includegraphics[width=\linewidth]{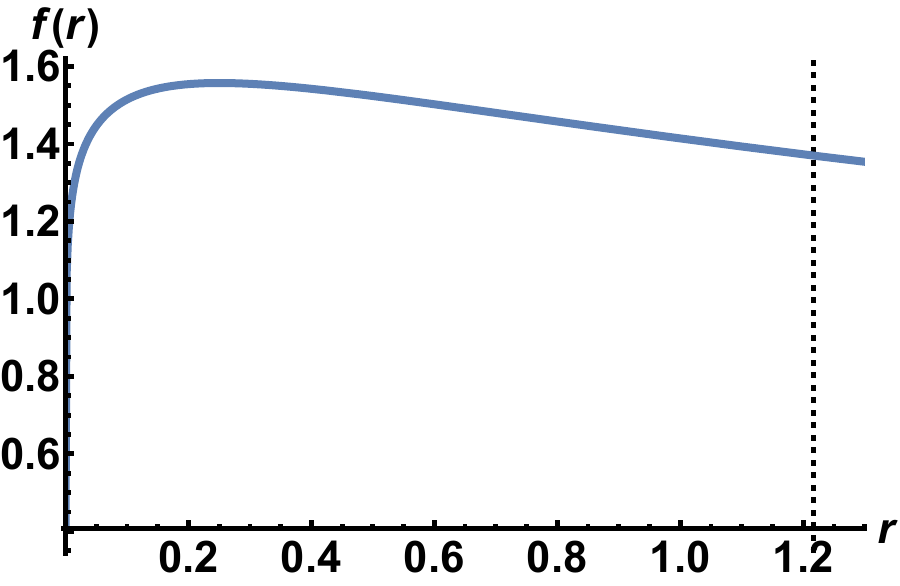}
\caption{$f$ solution}
  \end{subfigure}
  \begin{subfigure}[b]{0.326\linewidth}
    \includegraphics[width=\linewidth]{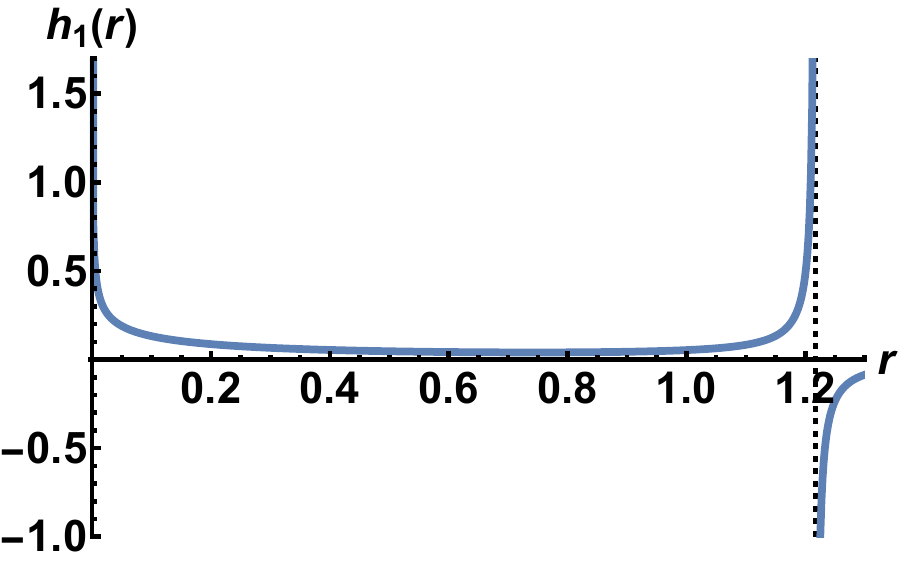}
\caption{$h_1$ solution}
  \end{subfigure}
  \begin{subfigure}[b]{0.326\linewidth}
    \includegraphics[width=\linewidth]{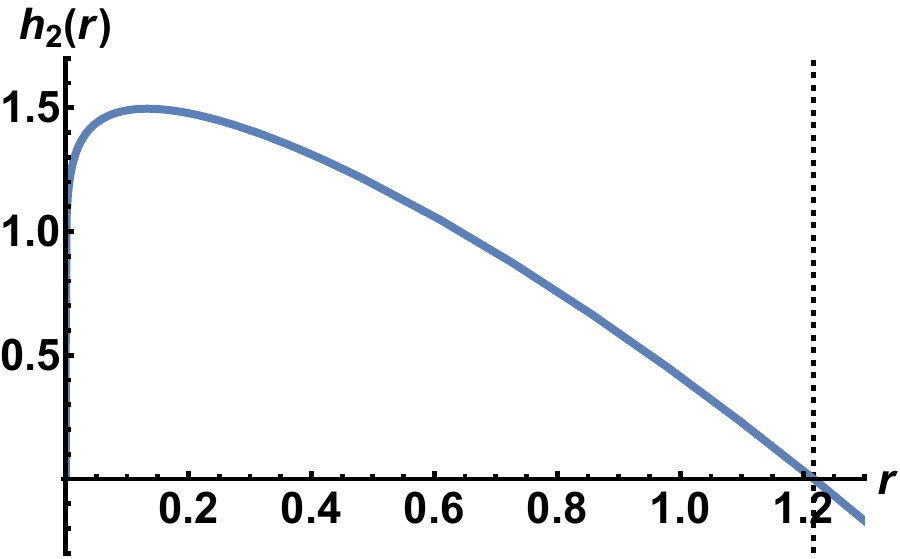}
\caption{$h_2$ solution}
  \end{subfigure}
  \caption{Numerical plots of the warp factors for the $SO(2)_{\text{diag}}$ symmetric solution from $SO(4,1)$ gauge group, evaluated at $s=1$, $B=2$, $c_1=\frac{1}{8}$, and $g=8$. The solution is regular within the interval $1<r<r_{2+}=1.217$. The vertical dashed line marks the boundary at $r=1.217$.}
  \label{Fig7}
\end{figure}

In $SO(4,1)$ gauge group with $\lambda=-1$, there exists only one possible range of $r$ for regular solutions. This is given by
\begin{equation}
0<r<r_{2+}\label{reg_range41}
\end{equation}
for $B>0$ and $s=+1$. An example of numerical plots of the warp factors for $B=2$, $c_1=\frac{1}{8}$ and $g=8$ is given in figure \ref{Fig7}. As $r\rightarrow0$, setting $r=3^{\frac{4}{3}}B^{\frac{2}{3}}c_1{}^{\frac{4}{3}}g^{\frac{4}{3}} \rho^{\frac{4}{3}}$, we find the seven-dimensional metric given in \eqref{SO(2)d_rto0_metric} near $\rho=0$. The seven-dimensional space-time is conformal to a product of $AdS_5$ and a cylinder as $r\rightarrow0$ as in the previous cases.

\begin{figure}[h!]
  \centering
    \includegraphics[width=0.5\linewidth]{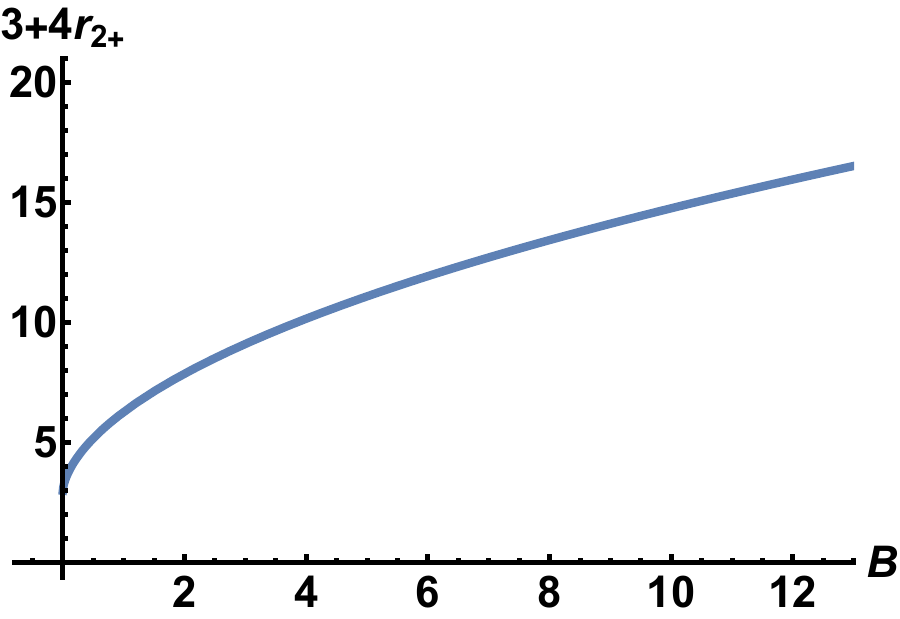}
  \caption{A numerical plot of the function $(3+4r_{2+})$ appearing in the condition \eqref{SO(4,1)_smooth_con} with $s=+1$ in $SO(4,1)$ gauge group. The function approaches $3$ as $B\rightarrow 0$.}
  \label{Fig8}
\end{figure}

As $r\rightarrow r_{2+}$, the seven-dimensional metric is approximately given by
\begin{equation}\label{SO(4,1)_orb_metric}
ds_7^2\approx\frac{64 B r_{2+}^{\frac{1}{10}}}{g^2 \sqrt{r_{2+}+1}}\left[ds^2_{AdS_5}+\frac{d\rho^2+c_1{}^2 g^2 \rho ^2 \left(3+4 r_{2+}\right)^2dz^2}{-W'\left(r_{2+}\right)\sqrt{r_{2+}}\left(r_{2+}+1\right)^{\frac{3}{2}}}\right]
\end{equation}
with the coordinate $\rho$ defined by $\rho=\frac{1}{2}\sqrt{r_{2+}-r}$. The $z$-circle shrinks smoothly giving rise to an $\mathbb{R}^2/\mathbb{Z}_l$ orbifold at $r=r_{2+}$ after imposing the condition
\begin{equation}\label{SO(4,1)_smooth_con}
c_1= \frac{1}{l g(3+4r_{2+})},\qquad l=1,2,3,\ldots\, .
\end{equation}
This function is always greater than $3$ in the regularity range \eqref{reg_range41} as shown in figure \ref{Fig8}.

Fixing the constant
\begin{equation}
q=-\frac{c_1 g}{2}\left(3+4 r_{2+}\right),
\end{equation}
we obtain the $SO(2)_{\text{diag}}$ gauge field
\begin{equation}
\mathbb{A}_d=4 c_1 \left(r_{2+}-r\right)
\end{equation}
which vanishes at $r=r_{2+}$. The Killing spinor at the endpoint $r=r_{2+}$ is explicitly given by
\begin{equation}\label{SO(4,1)_eta_Soln}
\eta=4i Y_0 e^{-\frac{i z }{2 l}} r_{2+}^{\frac{2}{5}}  \begin{pmatrix} 0 \\ 1 \end{pmatrix}.
\end{equation}

\subsection{Uplift to eleven dimensions}
We can now perform the uplift of the $SO(2)_{\textrm{diag}}$ symmetric solution in $SO(5)$ gauge group to eleven dimensions. All of the solutions given above can be uplifted by the same procedure as described in the previous section. We will mainly focus on the regular solution with $s=1$ in the range $0<r<r_{1-}$ since this leads to finite holographic central charge in the dual $N=1$ SCFT in four dimensions. 
\\
\indent In this case, we have 
\begin{equation}
\mathbb{A}_2=\mathbb{A}_1=2c_1(r_{1-}-r)
\end{equation} 
and 
\begin{equation}
\phi_2=\phi_1=-\frac{1}{10}\ln r\, . 
\end{equation}
With the same $S^4$ coordinates given in \eqref{S4_coor}, the eleven-dimensional metric becomes
\begin{eqnarray}
ds^2_{11}&=&\frac{2^6B\Delta^{\frac{1}{3}}r^{\frac{1}{10}}}{g^2\sqrt{1-r}}\left[ds^2_{AdS_5}+\frac{r^{-\frac{1}{2}}}{16W(1-r)^{\frac{3}{2}}}dr^2+\frac{g^2c_1^2W}{B}dz^2\right]\nonumber \\
& &+\frac{r^{\frac{1}{5}}\Delta^{-\frac{2}{3}}}{g^2}\left[(r^{-1}\cos^2\beta+\sin^2\beta)d\beta^2+\cos^2\beta\sin^2\alpha(d\varphi-g\mathbb{A}_1dz)^2\right.\nonumber \\
& &\left.+\cos^2\beta\cos^2\alpha(d\vartheta-g\mathbb{A}_1dz)^2\right]\label{11D_metric2}
\end{eqnarray}
with
\begin{equation}
\Delta=r^{-\frac{1}{5}}\cos^2\beta+r^{\frac{4}{5}}\sin^2\beta\, .
\end{equation}
\indent The four-form field strength reads
\begin{eqnarray}
F_{(4)}&=&-\frac{U}{4!g^3\Delta^2}\epsilon_{M_1M_2M_3M_4M_5}\mu^{M_1}D\mu^{M_2}\wedge D\mu^{M_3}\wedge D\mu^{M_4}\wedge D\mu^{M_5}\nonumber \\
& &+\frac{c_1r^{\frac{4}{5}}}{2g^2\Delta}\cos\beta\left[4(r^{-1}\cos^2\beta+\sin^2\beta)dr\wedge dz\wedge d\beta\wedge (\cos^2\alpha d\vartheta+\sin^2\alpha d\varphi)\right.\nonumber \\
& &\left.-\sin2\alpha\sin2\beta d\alpha \wedge dr\wedge dz \wedge(d\vartheta+d\varphi)\right]
\end{eqnarray}
with
\begin{equation}
U=(r^{\frac{8}{5}}-4r^{\frac{3}{5}})\sin^2\beta-(2r^{-\frac{2}{5}}+r^{\frac{3}{5}})\cos^2\beta\, .
\end{equation}
\indent The flux quantization proceeds along the same line as in the previous section. In particular, the integral of $F_{\alpha\beta\varphi\vartheta}$ gives the same result as in \eqref{quantize1}. On the other hand, the integral of $F_{rz\vartheta\beta}$ with $\alpha=0$ leads to the condition 
\begin{equation}
\frac{2r_{1-}}{\pi\ell_p^3g^3l(3-4r_{1-})}=\frac{2r_{1-}N}{l(3-4r_{1-})}=M\in \mathbb{Z}
\end{equation}
which gives
\begin{equation}
r_{1-}=\frac{3Ml}{2(N+2Ml)}\, .
\end{equation}
\indent We now compute the holographic central charge of the dual $N=1$ SCFT. With the eleven-dimensional metric of the form
\begin{equation}
ds^2_{11}=e^{2\mc{A}}\left(ds^2_{AdS_5}+ds^2_{M_6}\right),
\end{equation}
the holographic central charge is given by \cite{4D_holo_central_charge}
\begin{equation}
c=\frac{1}{2^7\pi^6\ell_p^9}\int e^{9\mc{A}}\textrm{vol}_{M_6}\, .
\end{equation}
By uplifting the metric ansazt \eqref{7Dmetrix} to eleven dimensions, we can identify
\begin{equation}
e^{\mc{A}}=\Delta^{\frac{1}{6}}\sqrt{f}\, .
\end{equation} 
The volume form on the internal manifold $M_6$ is given by
\begin{equation}
\textrm{vol}_{M_6}=\frac{\sqrt{h_1h_2}}{g^4f^3\Delta^{\frac{3}{2}}}\cos^3\beta\sin\alpha\cos\alpha dr\wedge dz \wedge d\beta \wedge d\alpha\wedge d\varphi\wedge d\vartheta\, .
\end{equation}
After performing the integrals along the angular coordinates, we find
\begin{equation}
c=\frac{1}{3(2\pi)^3\ell_p^9g^4}\int_0^{r_{1-}}\sqrt{f^3h_1h_2}dr\, .
\end{equation}
Using the explicit solutions for $f$, $h_1$ and $h_2$, we eventually obtain
\begin{equation}
c=\frac{2^{10}B^2r_{1-}}{3(\pi g^3\ell_p^3)^3l(1-r_{1-})(3-4r_{1-})}=\frac{3^2(2^6)N^2M^4l^3}{(N+2Ml)^3}
\end{equation}
in which we have used $B=r_{1-}^{\frac{3}{2}}\sqrt{1-r_{1-}}$ from the condition $W(r_{1-})=0$. For $M\sim N$, we find $c\sim N^3$.
\\
\indent We end this section by considering the uplifted metric near $r=0$ at which the seven-dimensional solution is singular as pointed out below \eqref{SO(2)d_rto0_metric}. In this limit, the metric \eqref{11D_metric2} becomes
\begin{eqnarray}
ds^2_{11}&\sim& r^{\frac{1}{30}}\left[ds^2_{AdS_5}+dz^2+r^{-\frac{1}{2}}dr^2\right]\nonumber \\
& &+r^{-\frac{2}{3}}\cos^2\beta\left\{d\beta^2+r\left[d\alpha^2+\sin^2\alpha D\varphi+\cos^2\alpha D\vartheta\right]\right\}
\end{eqnarray}
with $D\varphi=d\varphi-g\mathbb{A}_1 dz$ and $D\vartheta=d\vartheta-g\mathbb{A}_1dz$. Similar to the results of \cite{Bah_AD,Bah_M5,M5_disk_Couzen1,M5_disk_Couzen2}, we expect this limit to represent M5-branes smeared over one direction. In particular, as $r\rightarrow 0$, the four-form becomes
\begin{equation}  
F_{(4)}\sim \frac{1}{g^3}\cos\beta d\alpha \wedge d\beta\wedge D\vartheta\wedge D\varphi
\end{equation}
indicating the presence of M5-brane sources along $AdS_5\times S^1$. 
\section{$SO(2)\times SO(3)$ symmetric solutions}\label{Y_SO(2)_Sec}
\begin{figure}[h!]
  \centering
  \begin{subfigure}[b]{0.326\linewidth}
    \includegraphics[width=\linewidth]{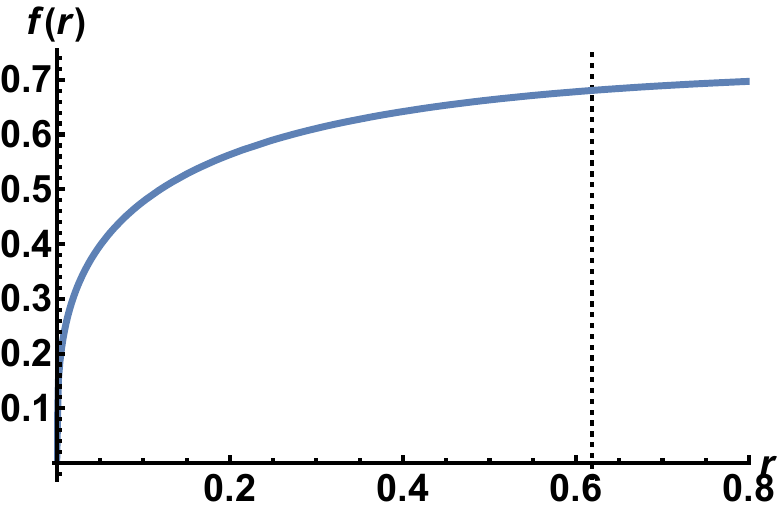}
\caption{$f$ solution}
  \end{subfigure}
  \begin{subfigure}[b]{0.326\linewidth}
    \includegraphics[width=\linewidth]{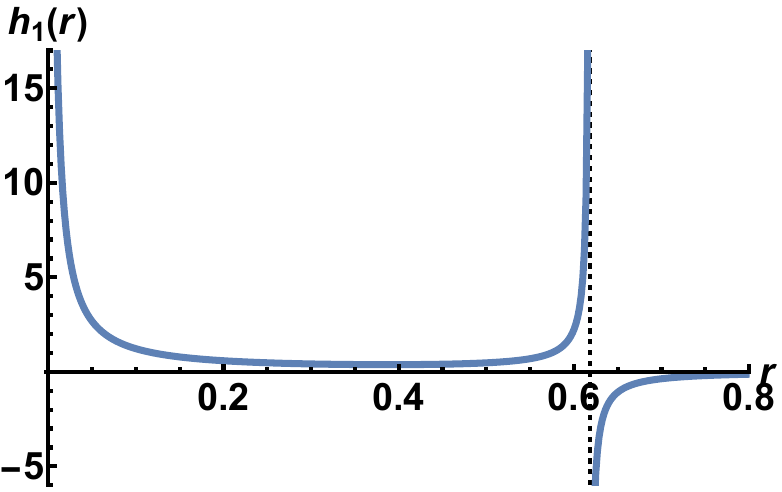}
\caption{$h_1$ solution}
  \end{subfigure}
  \begin{subfigure}[b]{0.326\linewidth}
    \includegraphics[width=\linewidth]{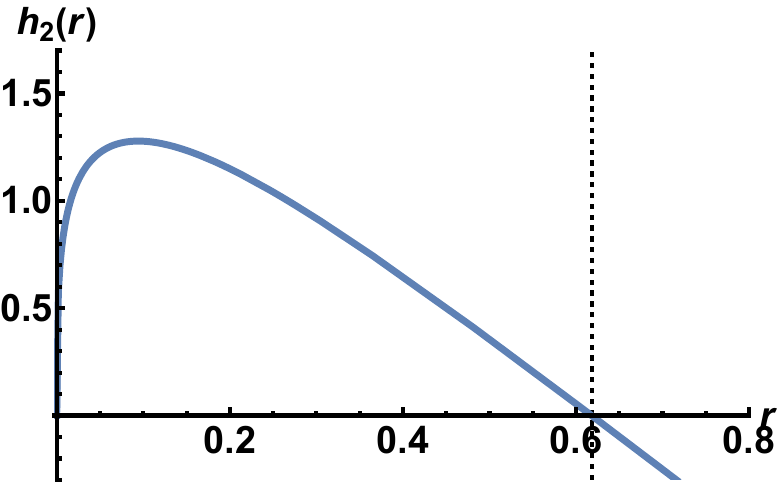}
\caption{$h_2$ solution}
  \end{subfigure}
  \caption{Numerical plots of the warp factors for the $SO(2)\times SO(3)$ symmetric solution from $SO(3,2)$ gauge group, evaluated at $s=1$, $B=1$, $c_1=\frac{1}{4}$, and $g=8$. The solution is regular within the interval $0<r<r_+=0.618$. The vertical dashed line marks the boundary at $r=r_+$.}
  \label{Fig13}
\end{figure}

We end the paper by considering another class of solutions with $SO(2)\times SO(3)$ symmetry. This can be found from the $SO(2)\times SO(2)$ truncation by setting $\mathbb{A}_2=0$ or equivalently setting the constant $a_2=0$. In this case, the condition \eqref{1st_Strongest_con} becomes
\begin{equation}
a_1 \lambda  e^{-2 (\phi_1+\phi_2)} \left(\kappa-\lambda e^{4 \phi_1+6\phi_2}\right)=0\label{SO(2)R_Strongest_con}
\end{equation}
which can be readily solved by
\begin{equation}
\phi_2=-\frac{2}{3}\phi_1+\frac{1}{6}\ln\left[\frac{\kappa}{\lambda}\right].
\end{equation}
This implies that $\frac{\kappa}{\lambda}$ must equal to $1$ in order to have real $\phi_1$ and $\phi_2$. Consequently, there are $AdS_5\times \Sigma$ solutions with $\phi_2=-\frac{2}{3}\phi_1$ only in $SO(5)$ ($\kappa=\lambda=1$) and $SO(3,2)$ ($\kappa=\lambda=-1$) gauge groups. We also note that setting $\phi_2=-\frac{2}{3}\phi_1$ leads to the $SL(5)/SO(5)$ coset representative of the form
\begin{equation}
\mc{V}=\textrm{diag}(e^{\phi_1},e^{\phi_1},e^{-\frac{2}{3}\phi_1},e^{-\frac{2}{3}\phi_1},e^{-\frac{2}{3}\phi_1})
\end{equation}
which implies the enhancement of symmetry from $SO(2)\times SO(2)$ to $SO(2)\times SO(3)$. Furthermore, as previously mentioned, with $\mathbb{A}_2=0$, supersymmetry is also enhanced to $16$ supercharges corresponding to $N=2$ superconformal symmetry in four dimensions. The solution in $SO(5)$ gauge group has already been studied extensively in \cite{Bah_AD,Bah_M5,M5_disk_Couzen1,M5_disk_Couzen2}, so we will mainly consider the solution from the non-compact $SO(3,2)$ gauge group. 
\\
\indent Setting $\kappa=\lambda$, $a_2=0$, and $\phi_2=-\frac{2}{3}\phi_1$, we can solve all the BPS equations as in the previous two sections and find the following solution
\begin{eqnarray}
f&=&\frac{64B e^{\frac{2 \phi_1}{3}}}{g^2\sqrt{s \left(e^{\frac{10 \phi_1}{3}}-\lambda \right)}},\\
h_1&=&\frac{400 B e^{4 \phi_1}(\phi'_1)^2}{9g^2 \left[s(e^{\frac{10 \phi_1}{3}}-\lambda) \right]^2 \left[B- e^{-\frac{10\phi_1}{3}} \sqrt{s \left(e^{\frac{10 \phi_1}{3}}-\lambda \right)}\right]},\\
h_2&=&\frac{64c_1{}^2 e^{\frac{2 \phi_1}{3}} \left[B- e^{-\frac{10\phi_1}{3}} \sqrt{s \left(e^{\frac{10 \phi_1}{3}}-\lambda \right)}\right]}{\sqrt{s \left(e^{\frac{10 \phi_1}{3}}-\lambda \right)}},\\
\widehat{\mathbb{A}}&=&c_1 \left(\lambda -2 e^{-\frac{10\phi_1}{3}}\right)=\mathbb{A}_1+\frac{2q}{g}\label{SO(2)R_A1_soln}
\end{eqnarray}
The Killing spinor is given by
\begin{equation}
\eta=\frac{2 \sqrt{2} Y_0 e^{-\frac{2 \phi_1}{3}+iqz}}{\left[s \left(e^{\frac{10\phi_1}{3}}-\lambda \right)\right]^{\frac{1}{8}}}\begin{pmatrix} \sqrt{ \sqrt{B} \lambda  e^{\frac{5 \phi_1}{3}}+s \left[s \left(e^{\frac{10 \phi_1}{3}}-\lambda \right)\right]^{\frac{1}{4}}} \\ -\sqrt{ \sqrt{B} \lambda  e^{\frac{5 \phi_1}{3}}-s \left[s \left(e^{\frac{10 \phi_1}{3}}-\lambda \right)\right]^{\frac{1}{4}}}\end{pmatrix}.\label{SO(2)_eta}
\end{equation}
It can also be verified that all of the BPS equations and the field equations are satisfied for
\begin{equation}\label{SO(2)R_sign_Con}
s \left(e^{\frac{10 \phi_1}{3}}-\lambda \right)>0\, .
\end{equation}
\indent At this point, it should be noted that this result is a subtruncation of the full $SO(2)$ invariant sector in which the residual $SO(2)$ symmetry is generated by $X_{12}$. Apart from the two $SO(2)\times SO(2)$ singlet scalars $\phi_1$ and $\phi_2$, there are additional $SO(2)$ singlet scalars. We have also considered the full $SO(2)$ invariant sector. However, it turns out that the field equations and BPS equations require that all these additional $SO(2)$ singlet scalars must vanish. Therefore, there is no supersymmetric $AdS_5\times \Sigma$ solution with only $SO(2)$ symmetry unbroken. In addition, it should be remarked that, with some appropriate identification, the solutions take the same form as the $SO(2)_{\text{diag}}$ symmetric solutions obtained from $N=2$ gauged supergravity coupled to three vector multiplets with $SO(4)$ and $SO(2,2)$ gauge groups studied in \cite{7D_N2_disk}. As in the previous cases, the higher-dimensional origin of these solutions in the context of half-maximal gauged supergravity is currently not known. Furthermore, in the framework of $N=2$ gauged supergravity, the solutions preserves only eight supercharges corresponding to $N=1$ superconformal symmetry in four dimensions.
\\
\indent With the function
\begin{equation}\label{Def_W_2}
W=B- e^{-\frac{10\phi_1}{3}} \sqrt{s \left(e^{\frac{10 \phi_1}{3}}-\lambda\right)},
\end{equation}
we find the seven-dimensional metric of the form
\begin{equation}
ds^2_7=\frac{64B\,e^{\frac{2 \phi_1}{3}}}{g^2\sqrt{s(e^{\frac{10\phi_1}{3}}-\lambda)}}\left[ds^2_{AdS_5}+\frac{25e^{\frac{10\phi_1}{3}}\,(\phi'_1)^2}{36W\left[s\left(e^{10\phi_1}-\lambda\right)\right]^{\frac{3}{2}}}dr^2+\frac{g^2c_1{}^2W}{B}dz^2\right].
\end{equation}
\indent To further analyize the solutions, we fix the solution of $\phi_1$ to be
\begin{equation}
\phi_1=-\frac{3}{10}\ln r\, .
\end{equation}
In terms of the radial coordinate, the seven-dimensional metric reads
\begin{equation}
ds^2_7=\frac{64Br^{\frac{3}{10}}}{g^2\sqrt{s(1-\lambda r)}}\left[ds^2_{AdS_5}+\frac{r^{-\frac{3}{2}}}{16W\left[s(1-\lambda r)\right]^{\frac{3}{2}}}dr^2+\frac{g^2c_1{}^2W}{B}dz^2\right]
\end{equation}
where
\begin{equation}
W=B-\sqrt{sr(1-\lambda r)}\, .
\end{equation}
\indent The scalar field $\phi_2$ and the non-vanishing $SO(2)$ gauge field are respectively given by
\begin{equation}\label{SO(2)diag_phi2_SO(4)_vector_Soln}
\phi_2=\frac{1}{5}\ln r\qquad\text{ and }\qquad \mathbb{A}_1=c_1 (\lambda -2 r)-\frac{2 q}{g}\, .
\end{equation}
\indent The equation $W=0$ admits two roots of the form
\begin{equation}\label{rpm_Roots}
r_{\pm}=\frac{s\pm \sqrt{1-4 B^2 \lambda  s}}{2 \lambda  s}\, .
\end{equation}
There are five possibilities for the regular solutions to exist. These are given by
\begin{eqnarray}
i)\qquad \lambda&=&+1,\quad s=+1,\quad 0<B<\frac{1}{2},\quad r\in(0,r_-)\cup (r_+,1),\quad\nonumber\\
ii)\qquad \lambda&=&+1,\quad s=+1,\quad B=\frac{1}{2},\quad r\in(0,r_\ast)\cup (r_\ast,1),\nonumber\\
iii)\qquad \lambda&=&+1,\quad s=+1,\quad B>\frac{1}{2},\quad r\in(0,1),\nonumber\\
iv)\qquad \lambda&=&+1,\quad s=-1,\quad B>0,\quad r\in(1,r_+)\nonumber\\
v)\qquad \lambda&=&-1,\quad s=+1,\quad B>0,\quad r\in(0,r_+)\nonumber
\end{eqnarray}
where $r_\ast=r_+=r_-=\frac{1}{2}$ when $B=\frac{1}{2}$ and $\lambda= s=1$. As previously mentioned, the case $\lambda=1$ has already been studied in \cite{Bah_M5} and \cite{7D_N2_disk}. Therefore, we will only consider the case of $\lambda=-1$ with an example of numerical solutions for the warp factors shown in figures \ref{Fig13}.

As $r\rightarrow0$, we can change to a new radial coordinate $R$ defined by $r=B^2c_1{}^4g^4 R^4$ and find the metric near $R=0$ of the form  
\begin{equation}\label{SO(2)_rto0_metric}
ds_7^2\approx\frac{64(B^4c_1{}^3 g^3 R^3){}^{2/5}}{g^2}\left[ds^2_{AdS_5}+c_1{}^2 g^2(dR^2+dz^2)\right].
\end{equation}
On the other hands, as $r\rightarrow r_+$, the seven-dimensional metric is approximately given by
\begin{equation}\label{SO(2)_spindle_metric}
ds_7^2\approx\frac{64B\,r_{+}^{\frac{3}{10}}}{g^2\sqrt{s(1+ r_{+})}}\left[ds^2_{AdS_5}+\frac{d\rho^2+c_1{}^2g^2\rho^2\left(1+2 r_{+}\right)^2dz^2}{-W'(r_{+})r_{+}^{\frac{3}{2}}\left[s\left(1+ r_{\pm}\right)\right]^{\frac{3}{2}}}\right]
\end{equation}
with the radial coordinate $\rho$ defined by $\rho=\frac{1}{2}\sqrt{r_+-r}$. The $z$ circle shrinks smoothly near the endpoint at $r= r_{+}$ if we impose
\begin{equation}\label{SO(2)diag_smooth_con}
c_1=\frac{1}{gl\left|1+2 r_{+}\right|}=\frac{1}{gl\sqrt{1+4B^2 s}},\qquad l=1,2,3,\ldots \, .
\end{equation}
In order for the $SO(2)$ gauge field to vanish at $r= r_{+}$, we fix the constant $q$ to be
\begin{equation}
q=-\frac{gc_1}{2}(1+2r_+)=-\frac{1}{2l}
\end{equation}
which gives 
\begin{equation}
\mathbb{A}_1=2c_1(r_+-r).
\end{equation}
The Killing spinor $\eta$ at $r= r_{+}$ takes the form
\begin{equation}\label{SO(2)_rpm_eta}
\eta=2 \sqrt{2} Y_0 e^{\frac{i \lambda z}{2 l}} r_{\pm}^{\frac{1}{5}}\begin{footnotesize}\begin{pmatrix} \sqrt{s+\lambda} \\ -\sqrt{s-\lambda} \end{pmatrix}\end{footnotesize}.
\end{equation}
Since the complete truncation ansatz for eleven-dimensional supergravity on $H^{3,2}$ is not known to date, we cannot perform the uplift of this solution to eleven dimensions. However, due to the absence of the limit with a decompactified $z$-circle, we expect to find a finite holographic central charge of the dual $N=2$ SCFT. 

\section{Conclusions and discussions}\label{conclusion}
We have studied supersymmetric $AdS_5\times \Sigma$ solutions for $\Sigma$ being a topological disk with a non-trivial $U(1)$ holonomy on the boundary from seven-dimensional maximal gauged supergravity with $CSO(p,q,5-p-q)$ gauge groups. We have found $SO(2)\times SO(2)$ symmetric solutions from $SO(5)$ and $SO(3,2)$ gauge groups as well as $SO(2)_{\text{diag}}$ symmetric solutions from $SO(5)$, $SO(4,1)$ gauge groups. All of these solutions preserve eight supercharges corresponding to $N=1$ superconformal symmetry in four dimensions. For solutions with $SO(2)\times SO(3)$ residual symmetry, apart from recovering the original solution obtained in \cite{Bah_M5} from $SO(5)$ gauge group, we have found a new $AdS_5\times \Sigma$ solution from $SO(3,2)$ gauge groups preserving sixteen supercharges. Although all of these solutions take the same form as those found in matter-coupled $N=2$ gauged supergravity studied in \cite{7D_N2_disk} after some appropriate identifications, the analogous solutions in the context of maximal gauged supergravity considered in the present paper have clear higher-dimensional origin in terms of a truncation of eleven-dimensional supergravity on $H^{p,5-p}$. 
\\
\indent For $SO(5)$ gauge group in which the complete truncation ansatz on $S^4$ is known, we have given the uplifted solutions to eleven dimensions. For $SO(2)_{\textrm{diag}}$ symmetric solution, we have also computed the holographic central charge of the dual $N=1$ SCFT in four dimensions. This $N=1$ SCFT arises from the near horizon geometry of M5-branes wrapped on a topological disk. In addition, all the remaining solutions in $SO(5)$ gauge group are asymptotic to the $AdS_7$ geometry at which the $z$-circle decompactifies rendering the holographic central charge to become infinite. In this case, the solutions should be interpreted as codimension-2 defects within the $N=(2,0)$ SCFT in six dimensions. The detailed holographic study of these solutions along the line of \cite{Gutperle_coD2_defect,Conti} also deserves further investigation.
\\
\indent For $SO(4,1)$ gauge group, we have found an $AdS_5\times \Sigma$ solution with $SO(2)_{\textrm{diag}}$ symmetry preverving eight supercharges, and for $SO(3,2)$ gauge group, we have found $AdS_5\times \Sigma$ solution preverving $SO(2)\times SO(3)$ symmetry and sixteen supercharges. Furthermore, the solutions are not asymptotic to the $AdS_7$ geometry with a decompactified $z$-circle. We then expect these solutions to be dual to $N=1$ and $N=2$ SCFTs in four dimentions, respectively. After some appropriate discrete quotient of $H^{4,1}$ and $H^{3,2}$ to have compact internal spaces, we could obtain finite holographic central charge of the dual $N=1$ and $N=2$ SCFTs due to the absence of a divergence at $r=1$. However, the $N=4$ gauged supergravities with $SO(4,1)$ and $SO(3,2)$ gauge groups do not admit a supersymmetric $AdS_7$ vacuum but half-supersymmetric domain walls. According to the DW/QFT correspondence, these domain wall vacua are dual to non-conformal $N=(2,0)$ theories in six dimensions. Therefore, unlike the solutions in $SO(5)$ gauge group, the aforementioned $N=1$ and $N=2$ four-dimensional SCFTs would arise from a compactification of these non-conformal $N=(2,0)$ theories on a topological disk. 
\\
\indent It would be interesting to construct explicit truncation ansatze for eleven-dimensional supergravity on $H^{p,q}\circ T^{5-p-q}$ and use the results to uplift solutions in $SO(4,1)$ and $SO(3,2)$ gauge groups. Identifying the dual $N=1$ and $N=2$ SCFTs dual to these solutions would also be of particular interest. Moreover, in obtaining $SO(2)\times SO(2)$ symmetric solutions, we have arbitrarily chosen a particular condition $a_2=-\kappa a_1$. Finding more general solutions without this condition would be worth considering and might lead to new solutions dual to four-dimensional $N=1$ SCFT with a $U(1)$ flavor symmetry. Finally, it could be interesting to find solutions describing non-conformal field theories in lower dimensions from 5-branes wrapped on $\Sigma$ in the present framework. This would lead to new solutions describing non-conformal 5-branes wrapped on a topological disk and generalize the results from $SO(5)$ $N=4$ gauged supergravity considered recently in \cite{Ferrero_non_con,Suh_non_con} to non-compact gauge groups. 
\appendix
\section{Bosonic field equations}\label{App_field_eqs}
For convenience, in this appendix, we collect all the bosonic field equations of the maximal gauged supergravity in seven dimensions. These equations are given by
\begin{eqnarray}
0&=&R_{\mu\nu}-\frac{1}{4}\mathcal{M}_{MP}\mathcal{M}_{NQ}(D_\mu\mathcal{M}^{MN})(D_\nu\mathcal{M}^{PQ})-\frac{2}{5}g_{\mu\nu}\mathbf{V}\nonumber\\&&-4\mathcal{M}_{MP}\mathcal{M}_{NQ}\left(\mathcal{H}_{\mu\rho}^{(2)MN}{{\mathcal{H}^{(2)PQ}}_{\nu}}^\rho-\frac{1}{10}g_{\mu\nu}\mathcal{H}_{\rho\sigma}^{(2)MN}\mathcal{H}^{(2)PQ\rho\sigma}\right)\nonumber\\
& &-\mathcal{M}^{MN}\left(\mathcal{H}_{\mu\rho\sigma M}^{(3)}{\mathcal{H}_\nu^{(3)\rho\sigma}}_N-\frac{2}{15}g_{\mu\nu}\mathcal{H}_{\rho\sigma\lambda M}^{(3)}{\mathcal{H}^{(3)\rho\sigma\lambda}}_N\right),
\end{eqnarray}
\begin{eqnarray}
0&=&D^\mu(\mathcal{M}_{MP}D_\mu\mathcal{M}^{PN})-\frac{g^2}{8}\mathcal{M}^{PQ}\mathcal{M}^{RN}\left(2Y_{RQ}Y_{PM}-Y_{PQ}Y_{RM}\right)\nonumber\\&&-\frac{4}{6}\mathcal{M}^{PN}\mathcal{H}_{\mu\nu\rho M}^{(3)}{\mathcal{H}^{(3)\mu\nu\rho}}_P-8\mathcal{M}_{MP}\mathcal{M}_{QR}\mathcal{H}_{\mu\nu}^{(2)PQ}
\mathcal{H}^{(2)RN\mu\nu}\nonumber\\&&+\frac{8}{5}\delta^N_M\left(\mathbf{V}+\mathcal{M}_{SP}\mathcal{M}_{QR}\mathcal{H}_{\mu\nu}^{(2)PQ}\mathcal{H}^{(2)RS\mu\nu }+\frac{1}{16}\mathcal{M}^{PQ}\mathcal{H}_{\mu\nu\rho P}^{(3)}{\mathcal{H}^{(3)\mu\nu\rho}}_Q\right)\nonumber\\&&+4g^2Z^{QT,P}Z^{NR,S}\mathcal{M}_{QM}(2\mathcal{M}_{TR}\mathcal{M}_{PS}-\mathcal{M}_{TP}\mathcal{M}_{RS})\nonumber \\
& &+4g^2Z^{QT,P}Z^{RS,N}\mathcal{M}_{QS}(2\mathcal{M}_{TP}\mathcal{M}_{RM}-\mathcal{M}_{TR}\mathcal{M}_{PM})\nonumber\\&&-4g^2\delta^N_MZ^{TU,P}Z^{QR,S}\mathcal{M}_{TQ}\left(\mathcal{M}_{UR}\mathcal{M}_{PS}
-\mathcal{M}_{UP}\mathcal{M}_{RS}\right) ,\label{H2_field_eqs}\\
0&=&4D_\nu(\mathcal{M}_{MP}\mathcal{M}_{NQ}\mathcal{H}^{(2)PQ\nu\mu })-\frac{g}{2}{X_{MNP}}^Q\mathcal{M}_{QR}D^\mu\mathcal{M}^{PR}\nonumber\\
& &-2\epsilon_{MNPQR}\mathcal{M}^{PS}{\mathcal{H}^{(3)\mu\nu\rho}}_S\mathcal{H}^{(2)QR}_{\nu\rho}+\frac{1}{9}e^{-1}\epsilon^{\mu\nu\rho\lambda\sigma\tau\kappa}\mathcal{H}^{(3)}_{\nu\rho\lambda M}\mathcal{H}^{(3)}_{\sigma\tau\kappa N},\\
0&=&D_\rho\left(\mathcal{M}^{MN}{\mathcal{H}^{(3)\rho\mu\nu}}_N\right)-2gZ^{NP,M}\mathcal{M}_{NQ}\mathcal{M}_{PR}\mathcal{H}^{(2)QR\mu\nu}\nonumber\\&&-\frac{1}{3}e^{-1}\epsilon^{\mu\nu\rho\lambda\sigma\tau\kappa}\mathcal{H}^{(2)MN}_{\rho\lambda}\mathcal{H}^{(3)}_{\sigma\tau\kappa N},\label{3formfieldEQ}\\
0&=&e^{-1}\epsilon^{\mu\nu\rho\lambda\sigma\tau\kappa}Y_{MN}\mathcal{H}^{(4)N}_{\lambda\sigma\tau\kappa}-6Y_{MN}\mathcal{M}^{NP}{\mathcal{H}^{(3)\mu\nu\rho}}_P\, .\label{fullSD}
\end{eqnarray}

\end{document}